\documentclass{JFM-FLM_Au}

\usepackage{bm}
\newcommand{\yo}[1]{{#1}}

\newcommand{\ii}{{\rm i}}

\lefttitle{Onuki and Venaille}
\righttitle{Journal of Fluid Mechanics}

\title{Disentangling discrete and continuous spectra of tidally forced internal waves in shear flow}

\author{Yohei Onuki\aff{1} \and Antoine Venaille\aff{2}}
\affiliation{\aff{1}Research Institute for Applied Mechanics, Kyushu University, Kasuga, Fukuoka 816-8580, Japan
\aff{2}CNRS, ENS de Lyon, Laboratoire de Physique, F-69342 Lyon, France}

\corresau{Yohei Onuki, onuki@riam.kyushu-u.ac.jp}

\begin{document}
\maketitle

\begin{abstract}
Generation of internal waves driven by barotropic tides over seafloor topography is a central issue in developing mixing and wave drag parameterizations for ocean circulation models. Traditional analytical approaches estimate the energy conversion rate from barotropic tides to internal waves using a modal expansion of the wave field. However, this framework becomes inadequate if a background shear flow is present, as singular solutions associated with critical levels emerge. To uncover the distinct roles of regular eigenmodes and singular solutions in tidal energy conversion, this study analytically investigates wave generation over a localized \yo{small topography} in the presence of shear flow \yo{without Coriolis force}. Applying horizontal Fourier and temporal Laplace transforms, we identify regions in the topographic wavenumber and forcing frequency space where unbounded energy growth occurs. These regions coincide with the spectrum of an operator governing free wave propagation and consist of discrete and continuous parts, which correspond to regular eigenmodes and singular solutions, respectively. Asymptotic evaluation of the Fourier integral reveals that the far-field response comprises standing wave trains linked to the discrete spectrum and evolving wave packets associated with the continuous spectrum. While the velocity amplitudes of the wave packets decay, their vertical velocity gradients grow during propagation, potentially leading to wave breaking. Finally, we derive a formula for the net barotropic-to-baroclinic energy conversion rate, extending the classical one by incorporating the contributions from both the discrete and continuous spectra.
\end{abstract}

\begin{keywords}

\end{keywords}

\section{Introduction}

Internal gravity waves generated by astronomical tides play pivotal roles in oceanic environments and the global climate system. These waves account for the majority of turbulent energy dissipation below the pycnocline, promoting vertical mixing of seawater\textemdash a process that significantly influences the distributions of heat, greenhouse gases, and nutrients, as well as the ocean’s meridional overturning circulation \citep{mackinnon2017climate,whalen2020internal}. Additionally, the momentum transfer associated with internal wave propagation is a crucial factor in shaping the slowly varying part of ocean currents \citep{shakespeare2019momentum}. 
Modeling the mixing and drag forces induced by internal waves has long been recognized as a key task in various areas of geosciences \citep{staquet2002internal}. Effectively parameterizing these effects in general circulation models remains a major challenge in modern physical oceanography \citep{fox2019challenges}.

For the generation of internal waves by the tidal forcing, the interaction between fluid motion and seafloor topography is essential. Planetary-scale variations in a tidal potential first induce a vertically homogeneous oscillatory horizontal motion of seawater, known as the barotropic tide. When this tide passes over sloping topography, it lifts the water masses, disturbing isopycnal surfaces to generate internal waves. These waves can have horizontal wavelengths exceeding 100~km, as observed by satellite altimeters \citep{zhao2016global,zhao2018global}. Modern numerical ocean models are capable of simulating the generation and propagation of large-scale internal waves with horizontal scales of approximately greater than 10~km \citep{niwa2014generation}. However, due to limitations in model resolution and uncertainty in wave decay processes \citep{onuki2018decay,olbers2020psi} on which the simulation result depends, analytical expressions are often used to estimate the energy generation rates of tidally forced internal waves in recent parameterizations \citep{de2019toward,de2020parameterization}. Therefore, precise theoretical formulations of internal wave generation under the forcing of barotropic tides are fundamental to developing a reliable parameterization.

Historically, numerous studies have analytically addressed the tidal generation of internal waves. A recent paper by \cite{papoutsellis2023internal}, as well as a comprehensive review by \cite{garrett2007internal}, \yo{provides a useful overview of the literature and} the relationships between various theoretical approaches. Most studies have examined internal wave generation in a static reference state. Under this framework, small-amplitude internal waves of a fixed frequency can be uniquely decomposed into discrete standing modes\textemdash a set of exact solutions for a linearized system over flat bottom topography derived by solving a Sturm-Liouville eigenvalue problem. This decomposition underpins key theoretical approaches such as the Green’s function method \citep{llewellyn2003tidal} and the coupled-mode system \citep{papoutsellis2023internal}, which address finite-height topography with small tidal excursions, and the formulation by \cite{khatiwala2003generation}, who considered small topography with finite tidal excursions. This strategy is effective even when incorporating steady barotropic flow \citep{dossmann2020asymmetric}.

The situation changes dramatically when a vertically sheared steady flow is introduced into the reference state. \yo{In the absence of planetary rotation, the vertical structure of a sheared internal gravity wave,
for a prescribed horizontal wave number and frequency, is governed by the Taylor--Goldstein equation. This equation admits not only regular eigenmodes but also singular solutions associated with critical levels, i.e., levels where the horizontal phase speed matches the background flow. As a result, the usual modal decomposition must be augmented; besides the discrete spectrum of regular modes, one must account for a continuous spectrum parameterized by the critical-level location.}

\yo{Continuous spectra and critical-layer singularities are a generic feature of wave-mean interaction in a wide range of geophysical and plasma settings. Here ``mean'' denotes a slowly varying background, such as a sheared mean flow in geophysical fluids and an inhomogeneous mean magnetic field in magnetohydrodynamics.} For a sheared stratified flow, \cite{booker1967critical}, \cite{engevik1971note}, and \cite{brown1980algebraic} made fundamental contributions \yo{to theoretical understanding of critical-layer processes}, as reviewed in relatively recent papers by \cite{roy2014linearized} and \cite{jose2015analytical}. Critical layer absorption of internal gravity waves is thought to be the primary driving source of the quasi-biennial oscillation in the stratosphere \citep{dunkerton1997role}. Critical layers also play key roles in the dissipation of atmospheric Rossby waves \citep{dickinson1970development,warn1976development,killworth1985rossby}, and in magnetohydrodynamics phenomena taking place in the outermost layer of the Earth's core or the solar tachocline \citep{nakashima2024two}. \yo{These studies highlight that singular solutions form essential parts in wave dynamics in various geophysical and astrophysical systems. By contrast, in the internal-tide literature, the continuous-spectrum contribution has received comparatively little attention, despite the ubiquity of shear in realistic ocean currents. Although several past studies \citep{hibiya1993control,lamb2018internal,masunaga2019strong}} considered periodic tidal forcing in a continuously sheared steady flow\footnote{Strictly speaking, \cite{hibiya1993control} considered a modulation of the velocity field induced by a spring-neap cycle of tidal mixing intensity, which should be distinguished from the tide-generated internal waves. However, their mathematical treatment \yo{falls in the same category.}}, \yo{they focused on the discrete-mode response and did not quantify the contributions of the continuous spectrum to the generated wave field.}

\yo{In addition to the spectral properties, velocity shear also complicates the energetics of the wave--mean-flow system. In a resting ocean, the barotropic-to-baroclinic conversion rate can be directly related to the radiated internal-wave energy flux \citep{petrelis2006tidal}. In the presence of vertical shear, however, internal waves exchange energy and momentum with the mean flow during propagation, so the wave energy alone does not form a closed budget. This motivates a reformulation of the disturbance energy equation that consistently accounts for mean-flow energy changes.

The purpose of the present study is to clarify the roles of background shear in tidally forced internal waves, with an eye toward mixing and wave-drag parameterizations. We focus on temporally periodic but spatially localized forcing by small-amplitude bottom topography, and analytically elucidate the resulting wave structure based on a spectral decomposition of the response into discrete and continuous constituents. We then formulate the net barotropic-to-baroclinic energy conversion rates separately for the discrete and continuous spectral components.

In this setting, the configuration considered closely matches the numerical simulations reported by \cite{lamb2018internal}. For clarity of interpretation and to enable direct comparison with this earlier work, we neglect the Coriolis effect. This simplification eliminates the additional singular behaviour that arises when the Doppler-shifted frequency approaches the Coriolis frequency \citep{jones1967propagation,xie2017interaction}, thereby making a fully analytical treatment tractable. In passing, recent works by \cite{le2025three} and \cite{maitland2025oceanic} have developed advanced numerical approaches to topographic internal-wave generation in shear flows including rotation; our analysis without rotation can therefore be viewed as a complementary step toward such more realistic configurations.

To formulate energy conversion rates while making the wave--mean-flow interaction explicit, we use standard wave-activity diagnostics---pseudomomentum and its associated pseudoenergy---which relate the induced mean-flow response to quadratic forms of the leading-order wave fields. These quantities arise naturally from the symmetry properties of the governing equations and provide conserved measures in the absence of forcing and dissipation (e.g., \citealt{shepherd1990symmetries}; \citealt{buhler2014waves}). They remain well defined even in the presence of rotation and therefore provide a natural starting point for extensions to more general oceanic conditions.

With this framework in place, we begin by deriving the leading-order linear system and boundary conditions in the small-topography regime and by specifying the asymptotic expansion that underpins the remainder of the analysis (Section~\ref{sec:problem_setup}). We then recast the linear problem using horizontal Fourier and temporal Laplace transforms, so that the forced response is controlled by the resolvent of a wave operator (Section~\ref{sec:general_solution}). This formulation makes the spectral content of the dynamics explicit: isolated singularities correspond to regular eigenmodes (discrete spectrum), while critical-level singular solutions generate branch-cut contributions that form a continuous spectrum. Identifying where a forcing frequency intersects these spectra provides a precise criterion for resonant behaviour and sets the stage for a systematic decomposition of the response.

Building on this spectral viewpoint, we solve the stationary boundary-value problem for each tidal harmonic and connect the Fourier-space solution to physical-space wave fields for localized forcing (Section~\ref{sec:stationary_solution}). An asymptotic evaluation of the inverse Fourier integral yields a transparent far-field picture: discrete poles produce persistent standing-mode wave trains, whereas branch cuts associated with the continuous spectrum produce dispersive wave packets whose velocity amplitudes decay algebraically while their vertical gradients grow during propagation. Having established this structure, we turn to energetics and derive a closed expression for the net barotropic-to-baroclinic conversion rate in the presence of shear, using pseudomomentum and pseudoenergy to diagnose the induced mean-flow response and to partition the conversion into discrete- and continuous-spectrum contributions (Section~\ref{sec:energetics}). We conclude by discussing implications and limitations of the idealised setting and outlining extensions toward more general oceanic configurations, including rotation (Sections~\ref{sec:discussion} and \ref{sec:conclusions}).
}

\section{Problem setup: Boussinesq fluid with rigid bottom and upper boundaries} \label{sec:problem_setup}
\yo{
This section sets up the mathematical model used throughout the paper. We consider a two-dimensional inviscid, incompressible fluid under the Boussinesq approximation, in which a prescribed barotropic tidal flow is superimposed on a steady vertically sheared current over a corrugated bottom. Their interaction with a small-amplitude topography excites internal gravity waves that propagate through a stably stratified background with shear. In what follows, we derive the linear disturbance equations governing this wave response. The sole approximation invoked is the small-topography assumption, which enables a systematic asymptotic expansion and linearization of the governing equations; in particular, we do not introduce any additional scale separation in space or time.
}

\subsection{Governing equations}
\yo{We start from the incompressible Euler equations under the Boussinesq approximation.} The governing equations are
\begin{subequations} \label{eq:equation_boussinesq}
\begin{align}
\frac{\partial \bm{u}}{\partial t} + \bm{u} \cdot \nabla \bm{u} & = - \frac{\nabla p + g \rho \bm{e}_z}{\rho^\star} \label{eq:equation_boussinesq_u} \\
\frac{\partial \rho}{\partial t} + \bm{u} \cdot \nabla \rho & = 0 \label{eq:density_equation} \\
\nabla \cdot \bm{u} & = 0 \label{eq:equation_boussinesq_continuity} ,
\end{align}
\end{subequations}
where $\bm{u}(x,z,t) = (u, w)$ is the velocity vector, $p(x,z,t)$ is the pressure, $\rho(x,z,t)$ is the density, $\rho^\star$ is the reference density, $g$ is the gravitational acceleration, and $\bm{e}_z$ is the upward unit vector. For the sake of conciseness, differentiation is also denoted as $\partial u / \partial t = \partial_t u = u_t$ in the following. Fluid is bounded in the vertical direction by a flat top ($z = H$) and a corrugated bottom ($z = h(x)$) boundaries, such that
\begin{align} \label{eq:bottom_condition_1}
w (x, H, t) = 0, \quad w (x, h, t) = h_x u(x, h, t) 
\end{align}
are understood. Although the equation for the density \eqref{eq:density_equation} does not \yo{require boundary conditions in general}, to simplify the problem, we shall assume that density is homogeneous along each boundary such that $\rho = \rho_{min}$ at $z = H$ and $\rho = \rho_{max}$ at $z = h(x)$ hold, where $\rho_{min}$ and $\rho_{max}$ are the minimum and maximum values of the density in the system, respectively. If this condition is initially fulfilled, it remains valid all the time.

\begin{figure}
\centerline{\includegraphics[width=0.5\columnwidth]{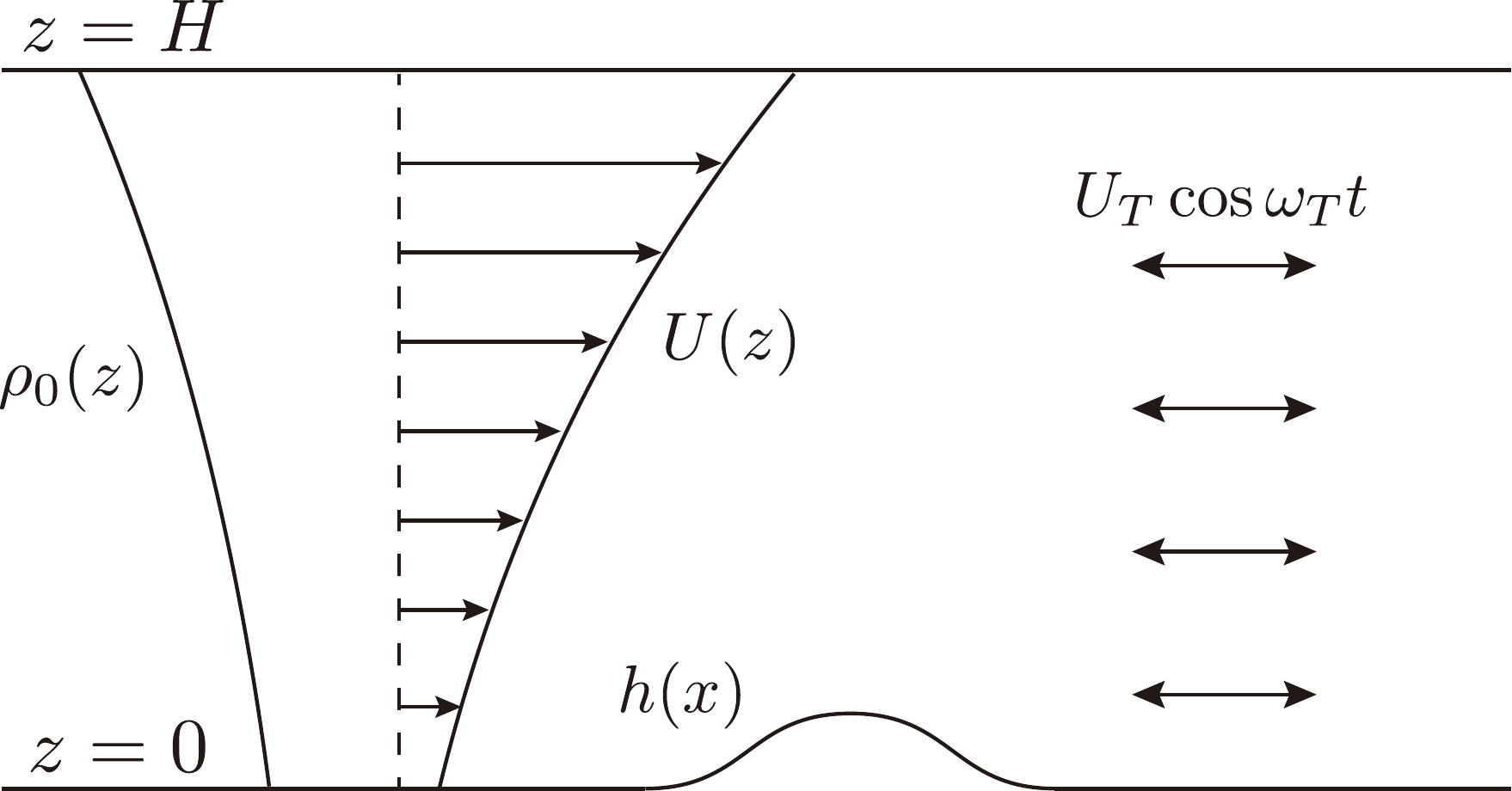}}
\caption{The situation considered in this study. Internal gravity waves are generated over a \yo{spatially localized small bottom topography} forced by an oscillatory barotropic flow in the presence of a sheared background current.}
\label{fig:concept}
\end{figure}

We define the reference state as a combination of density stratification, vertically sheared horizontal flow, and homogeneous oscillating tidal flow; i.e., $\rho = \rho_0 (z)$, $u = u_0(z, t) \equiv U(z) - U_T \cos \omega_T t$ and $w = 0$. The stratification is statically stable, and the buoyancy frequency is defined as $N = \sqrt{(- g / \rho^\star) d \rho_0 / d z}$. Figure~\ref{fig:concept} illustrates the situation under consideration.

For a flat bottom boundary case, $h = 0$, the reference state $(\bm{u}, \rho) = (\bm{u}_0, \rho_0)$ is an exact solution of (\ref{eq:equation_boussinesq}), but generally the bottom topography acts as an obstacle for the current, and a flow rising up the sloping boundary generates internal gravity waves. Therefore, we may write the solution as a superposition of the reference components and wave disturbances,
\begin{align}
u = u_0 + u', \quad w = w', \quad \rho = \rho_0 + \rho', \quad p = p_0 + p' .
\end{align}
Inserting these expressions into (\ref{eq:equation_boussinesq}) and (\ref{eq:bottom_condition_1}), and regarding the reference pressure as
\begin{align}
p_0 = g \int^H_z \rho_0 (\zeta) d\zeta - \rho^\star \omega_T U_T x \sin\omega_T t ,
\end{align}
we derive a set of governing equations for the wave motion,
\begin{subequations} \label{eq:equation_disturbance}
\begin{align}
\frac{\partial \bm{u}'}{\partial t} + u_0 \frac{\partial \bm{u}'}{\partial x} + u_{0z} w' \bm{e}_x + \bm{u}' \cdot \nabla \bm{u}' & = - \frac{\nabla p' + g \rho' \bm{e}_z}{\rho^\star} \label{eq:equation_u_prime} \\
\frac{\partial \rho'}{\partial t} + u_0 \frac{\partial \rho'}{\partial x} - \frac{\rho^\star N^2}{g} w' + \bm{u}' \cdot \nabla \rho' & = 0 \\
\nabla \cdot \bm{u}' & = 0 ,
\end{align}
\end{subequations}
as well as their boundary conditions,
\begin{align}
w' (x, H, t) = 0, \quad w' (x, h, t) = h_x \left[ u_0(h, t) + u'(x, h, t) \right] . \label{eq:bottom_condition_2}
\end{align}
In the following, we shall assume that the bottom topography and the disturbance at the initial time are horizontally localized such that $\lim_{x \to \pm \infty} \{ h, u', w' \} = 0$ at $t = 0$ is understood. This condition, as well as the incompressible constraint \eqref{eq:equation_boussinesq_continuity}, makes the net horizontal volume transport governed by the reference state and \yo{independent of} the bottom-generated disturbance. This point is made explicit by a formula, $\int_{h(x)}^H u(x, z, t) dz = \int_0^H u_0(z, t) dz$, or equivalently
\begin{align} \label{eq:volume_transport}
\int^H_h u' dz = \int_0^h u_0 dz ,
\end{align}
whose horizontal derivative conforms to the bottom boundary condition in (\ref{eq:bottom_condition_2}).

\subsection{Scaling and \yo{dimensionless constants}} \label{sec:scaling}
Up to now, we have not made any approximations; the set of equations governing the disturbance motions, (\ref{eq:equation_disturbance}) and (\ref{eq:bottom_condition_2}), is equivalent to the original ones, (\ref{eq:equation_boussinesq}) and (\ref{eq:bottom_condition_1}). A key step to making the problem more tractable is linearizing the equations based on proper scaling and asymptotic expansion. We write the typical velocity as $U^\star$, the typical buoyancy frequency as $N^\star$, and the typical topographic height as $h^\star$. We then scale and redefine the variables as
\begin{gather*}
x + \int^t_0 U_T \cos \omega_T t' dt' = H \tilde{x}, \quad z = H \tilde{z}, \quad t = \frac{\tilde{t}}{N^\star} , \\
(u', w') = U^\star \left(- \frac{\partial \tilde{\psi}}{\partial \tilde{z}}, \frac{\partial \tilde{\psi}}{\partial \tilde{x}} \right), \quad \rho' = \frac{\rho^\star U^\star N^\star \tilde{\rho}}{g}, \quad p' = \rho^\star U^\star N^\star H \tilde{p} , \\
\left( U, U_T \right) = U^\star \left( \tilde{U}, \tilde{U}_T \right), \quad \omega_T = N^\star \tilde{\omega}_T, \quad N = N^\star \tilde{N}, \quad h = h^\star \tilde{h} ,
\end{gather*}
where the horizontal coordinate is changed to follow the background oscillating current, and $\tilde{\psi}$ and $\tilde{\rho}$ represent the streamfunction and the density perturbation for the wave component, respectively.

In the present setting, two dimensionless constants naturally arise as
\begin{align}
\epsilon \equiv \frac{N^\star h^\star}{U^\star}, \quad \mu \equiv \frac{U^\star}{N^\star H} ,
\end{align}
in which $\epsilon$ measures the nonlinearity of the topographically generated internal waves, and $\mu$ represents the Froude number of the reference flow. 

\subsection{\yo{Asymptotic regime and expansion}}
\yo{To apply the asymptotic expansion, we assume a small topography 
regime $\epsilon \ll 1$ so that the nonlinear effects are negligible in the leading order.} If we regard the full depth $H$ as the typical length scale of the reference velocity, $\mu^{-2}$ roughly represents the gradient Richardson number. Accordingly, the assumption that $\mu$ is small would allow the use of the WKB approximation in the vertical direction. However, we do not rely on such \yo{an approximation} here; from now on, we set $\mu=1$, which means that the velocity variables are scaled by $N^\star H$. For simplicity, we will omit the tilde on all symbols below.

We expand the unknown variables in terms of \yo{a small parameter} $\epsilon$ as
\begin{align}
(\psi, \rho, p) = \epsilon (\psi_1, \rho_1, p_1) + \epsilon^2 (\psi_2, \rho_2, p_2) + \ldots . 
\end{align}
Accordingly, from (\ref{eq:equation_disturbance}), (\ref{eq:bottom_condition_2}) and (\ref{eq:volume_transport}), the equations governing the first-order terms are derived as
\begin{subequations} \label{eq:first_order_equation}
\begin{align}
\frac{\partial \nabla^2 \psi_1}{\partial t} & = - \frac{\partial \rho_1}{\partial x} - U \frac{\partial \nabla^2 \psi_1}{\partial x} + U_{zz} \frac{\partial \psi_1}{\partial x}  \label{eq:first_order_equation_psi} \\
\frac{\partial \rho_1}{\partial t} & = N^2 \frac{\partial \psi_1}{\partial x} - U \frac{\partial \rho_1}{\partial x}  \label{eq:first_order_equation_rho}
\end{align}
\end{subequations}
with the boundary conditions,
\begin{align} \label{eq:first_order_boundary_condtion}
\psi_1 (x, 1, t) = 0, \quad \psi_1 (x, 0, t) = \left[ U(0) - U_T \cos \omega_T t \right] h \left( x - \int^t_0 U_T \cos \omega_T t' dt' \right) .
\end{align}
Note that the bottom boundary condition is now set at $z = 0$ instead of $z = \epsilon h$. This treatment is common for topographic wave generation theory and valid as far as we consider linear processes. However, it requires some modifications in the energy budget equation. \yo{We will discuss this point in Section~\ref{sec:energetics}}.

Compared to the original nonlinear equations, the reduced linear equations, \eqref{eq:first_order_equation} and \eqref{eq:first_order_boundary_condtion}, are simple enough to be solved analytically. In the following, we assume $N(z)$ and $U(z)$ are sufficiently smooth functions and satisfy the condition for the Richardson number $N^2 / U_z^2 > 1/4$ everywhere in $0 \leq z \leq 1$, which ensures the flow stability \citep{miles1961stability,howard1961note}. Besides, we also assume $U_z>0$ to let $N^2 / U_z^2$ be finite.

\section{\yo{Formal solutions of the linear problem}} \label{sec:general_solution}

\yo{The linearized system \eqref{eq:first_order_equation}--\eqref{eq:first_order_boundary_condtion} describes the leading-order internal-wave response generated by small topography, while neglecting the feedback on the prescribed background flow. In this section, we obtain a formal solution of this initial-value problem using a horizontal Fourier transform and a temporal Laplace transform. This formulation expresses the response in terms of the resolvent of a wave operator.

Our main goal is to make the spectral structure of this operator explicit. This spectral viewpoint underpins the more explicit stationary solutions derived in Section~\ref{sec:stationary_solution} and provides the foundation for the energetics and conversion-rate formulas developed in Section~\ref{sec:energetics}.
}

\subsection{\yo{Horizontal Fourier transform}}
Since the coefficients in the equations \eqref{eq:first_order_equation} do not depend on $x$, it is useful to take the Fourier transform in the horizontal direction,
\begin{align} \label{eq:Fourier_transform}
\left[ \psi_1(x,z,t), \rho_1(x,z,t), h(x) \right] = \frac{1}{2\pi} \int^\infty_{-\infty} \left[ \hat{\psi}_1(k,z,t), \hat{\rho}_1(k,z,t), \hat{h}(k) \right] e^{\ii kx} dk .
\end{align}
Following \cite{bell1975lee}, we apply the Jacobi-Anger expansion to the Fourier transform of the bottom boundary condition in \eqref{eq:first_order_boundary_condtion} in terms of the harmonics of the tidal frequency to obtain
\begin{align} \label{eq:boundary_Bessel}
\hat{\psi}_1 (k, 0, t) = \sum_{n = -\infty}^\infty \frac{(k U_0 - \omega_n) \hat{h}}{k} J_n \left( \frac{k U_T}{\omega_T} \right) e^{- \ii \omega_n t} ,
\end{align}
where $U_0 \equiv U(0)$, $\omega_n \equiv n \omega_T$, and $J_n$ is the $n$th-order Bessel function of the first kind\footnote{\yo{Equation~\eqref{eq:boundary_Bessel}} involves an infinite number of terms corresponding to the harmonics of the tidal frequency. Their relative contributions depend on the parameters $k U_T / \omega_T$ and $k U_0 / \omega_T$. If the steady current is absent at the bottom ($U_0 = 0$), it is classically known that the tidal frequency $\vert n \vert = 1$ is dominant when $k U_T / \omega_T \ll 1$, while high harmonics, $\vert n \vert > 1$, are important when $k U_T / \omega_T \gtrsim 1$. On the other hand, if we take into account the steady current, $U_0 \neq 0$, a stationary component, $n = 0$, also appears \citep{shakespeare2020interdependence} and is dominant when $k U_T / \omega_T \ll 1$.}. The governing equations \eqref{eq:first_order_equation} are transformed into
\begin{subequations} \label{eq:equation_psi_k}
\begin{align}
\frac{\partial \nabla_k^2 \hat{\psi}_1}{\partial t} & = - \ii k \hat{\rho}_1 - \ii k U \nabla_k^2 \hat{\psi}_1 + \ii k U_{zz} \hat{\psi}_1 \\
\frac{\partial \hat{\rho}_1}{\partial t} & = \ii k N^2 \hat{\psi}_1 - \ii k U \hat{\rho}_1 \label{eq:equation_rho_k}
\end{align}
\end{subequations}
with $\nabla^2_k \equiv -k^2 + \partial_z^2$. Equations \eqref{eq:boundary_Bessel} and \eqref{eq:equation_psi_k} as well as the upper boundary condition, \yo{$\hat{\psi}_1(k, 1, t) = 0$}, redefine the problem to be solved. Classically, there are two strategies to address this type of problem. The first is to directly find the solution that satisfies the inhomogeneous boundary condition. The second is to decompose the solution into two parts: one that satisfies the inhomogeneous boundary condition and is easily detectable, and the other that satisfies homogeneous boundary conditions and can be solved separately. This section employs the second approach \yo{because it provides a better insight into the basic spectral properties of waves inherent in shear flow problems.} The first strategy will be applied in Section~\ref{sec:stationary_solution}, where we discuss the stationary response solution.

\subsection{\yo{Potential and vortical flow decomposition}}

\yo{To handle the inhomogeneous boundary condition at the bottom, we employ the Helmholtz decomposition for the velocity field to write the streamfunction as $\hat{\psi}_1 = \hat{\psi}_1^p + \hat{\psi}_1^v$. Here, $\hat{\psi}_1^p$ and $\hat{\psi}_1^v$ represent the potential and vortical flow parts, respectively.} The potential flow solves $\nabla^2_k \hat{\psi}_1^p = 0$ and satisfies the inhomogeneous bottom boundary condition \eqref{eq:boundary_Bessel} and the homogeneous upper boundary condition. An analytical solution is immediately derived as
\begin{align}
\hat{\psi}^p_1 = \frac{\sinh \left[ k(1 - z) \right]}{\sinh k} \sum_{n = -\infty}^\infty \frac{(k U_0 - \omega_n) \hat{h}}{k} J_n \left( \frac{k U_T}{\omega_T} \right) e^{- \ii \omega_n t} .
\end{align}
The vertical structure of the potential flow depends on the magnitude of the horizontal wavenumber. When the wavenumber is small, $\vert k \vert \ll 1$, the horizontal component of the potential flow, $-\partial_z \hat{\psi}^p_1$, is homogeneous in the vertical direction, representing the barotropic response in the hydrostatic regime. If the wavenumber is large, $\vert k \vert \gg 1$, the potential flow is localized close to the bottom and exponentially weak in the upper part of the fluid.

Now that a solution of the potential flow part is obtained, the remaining vortical flow part $\hat{\psi}_1^v$ as well as the density perturbation $\hat{\rho}_1$ are determined from \eqref{eq:equation_psi_k} as the solutions of a set of inhomogeneous equations,
\begin{subequations} \label{eq:equation_psiv_k}
\begin{align}
\frac{\partial \nabla_k^2 \hat{\psi}^v_1}{\partial t} & = - \ii k \hat{\rho}_1 - \ii k U \nabla_k^2 \hat{\psi}^v_1 + \ii k U_{zz} \hat{\psi}^v_1 + \ii k U_{zz} \hat{\psi}^p_1 \\
\frac{\partial \hat{\rho}_1}{\partial t} & = \ii k N^2 \hat{\psi}^v_1 - \ii k U \hat{\rho}_1 + \ii k N^2 \hat{\psi}^p_1 ,
\end{align}
\end{subequations}
with the homogeneous boundary conditions, $\hat{\psi}^v_1 (k, 1, t) = \hat{\psi}^v_1 (k, 0, t) = 0$. Our present problem is thus to investigate the linear response of prognostic variables $(\hat{\psi}^v_1, \hat{\rho}_1)$ to time-dependent external forcing originating from $\hat{\psi}_1^p$ as governed by \eqref{eq:equation_psiv_k}. Equations \eqref{eq:equation_psiv_k} are valid for each $k$ independently. Since all the terms on the right-hand sides vanish when $k=0$, we may set $\hat{\psi}_1^v(0, z, t) = \hat{\rho}_1(0, z, t) = 0$.

\subsection{\yo{Formal solutions derived from Laplace transform}} \label{sec:formal_solution}

To derive the general solution of \eqref{eq:equation_psiv_k}, it is convenient to \yo{introduce a linear operator $\mathsfbi{M}_k$ and rewrite the system as a forced Schr\"odinger-type evolution equation}
\begin{align} \label{eq:equation_Mk}
\ii \frac{\partial \hat{\bm{v}}}{\partial t} = \mathsfbi{M}_k \hat{\bm{v}} + \sum_{n = -\infty}^\infty \hat{\bm{f}}_n e^{- \ii \omega_n t} ,
\end{align}
where
\begin{align} \label{eq:vector_and_forcing}
\hat{\bm{v}}(k, z, t) = \left( \begin{array}{c}
\nabla_k^2 \hat{\psi}^v_1 \\
\hat{\rho}_1
\end{array} \right), \quad
\hat{\bm{f}}_n (k, z) = - \frac{(k U_0 - \omega_n) \hat{h} \sinh \left[ k(1 - z)\right]}{\sinh k} J_n \left( \frac{k U_T}{\omega_T} \right) \left( \begin{array}{c}
U_{zz} \\
N^2
\end{array} \right)
\end{align}
and
\begin{align} \label{eq:operator_Mk}
\mathsfbi{M}_k = \left( \begin{array}{cc}
  k U - k U_{zz} \nabla_k^{-2} & k \\
  - k N^2 \nabla^{-2}_k & k U
\end{array} \right) .
\end{align}
Here, the inverse of the Laplacian is defined \yo{by}
$\nabla_k^{-2} A = B \Leftrightarrow A = \nabla_k^2 B$ with $\left. B \right|_{z = 0,1} = 0$ for a \yo{prescribed $A$, and its explicit representation is given} in \eqref{eq:inverse_laplacian}. \yo{For an arbitrary initial condition $\hat{\bm{v}}(k,z,0) = \hat{\bm{v}}_0(k, z)$, the solution of \eqref{eq:equation_Mk} can be formally obtained using a temporal Laplace transform. We define}
\begin{align} \label{eq:laplace_transform}
\tilde{\bm{v}} (k, z; \sigma) = \int_0^{\infty} \hat{\bm{v}} (k, z, t) e^{\ii \sigma t} d t ,
\end{align}
where $\sigma$ is a complex number \yo{with $\Im \sigma$ sufficiently large for the integral to converge. Applying this transform to \eqref{eq:equation_Mk} yields}
\begin{align} \label{eq:equation_laplace_transformed}
\sigma \tilde{\bm{v}} = \mathsfbi{M}_k \tilde{\bm{v}} + \ii \hat{\bm{v}}_0 + \sum_{n = -\infty}^\infty \frac{\ii \hat{\bm{f}}_n}{(\sigma - \omega_n)} .
\end{align}
The inverse transform \yo{then gives the general solution}
\begin{align} \label{eq:general_solution}
\hat{\bm{v}} = \frac{\ii}{2 \pi} \int_{-\infty + \ii \gamma}^{\infty + \ii \gamma} \frac{e^{- \ii \sigma t} \hat{\bm{v}}_0}{\sigma \mathsfbi{I} - \mathsfbi{M}_k} d\sigma + \frac{\ii}{2 \pi} \sum_{n = - \infty}^\infty \int_{-\infty + \ii \gamma}^{\infty + \ii \gamma} \frac{e^{- \ii \sigma t} \hat{\bm{f}}_n}{(\sigma - \omega_n) (\sigma \mathsfbi{I} - \mathsfbi{M}_k)} d\sigma ,
\end{align}
where $\mathsfbi{I}$ is the identity matrix\yo{, and $\gamma > 0$ is chosen so that \eqref{eq:laplace_transform} converges for all $\sigma$ with $\Im\,\sigma > \gamma$.}

\yo{Equation \eqref{eq:general_solution} expresses the solution as an inverse Laplace transform whose integrand contains $(\sigma \mathsfbi{I}-\mathsfbi{M}_k)^{-1}$, the resolvent of the operator $\mathsfbi{M}_k$. In general, such inverse transforms are evaluated by deforming the contour and accounting for the singular behaviour of the integrand in the complex $\sigma$-plane. It is therefore essential to identify the set of values of $\sigma$ for which the resolvent fails to be well defined. For systems with finitely many degrees of freedom, the relevant singularities typically reduce to isolated poles, so the residue theorem yields the familiar decomposition into free modal oscillations and a particular forced response. In the present shear-flow problem, however, the unknowns are functions of the continuous vertical coordinate $z$, and the resolvent can exhibit non-isolated singular behaviour that cannot be captured by residues alone. We refer to this set as the spectrum of $\mathsfbi{M}_k$, which we determine in the next subsection.}

\subsection{\yo{Spectrum of the operator $\mathsfbi{M}_k$}} \label{sec:spectrum}

\yo{
To make the above statement concrete, we regard the resolvent $(\sigma \mathsfbi{I}-\mathsfbi{M}_k)^{-1}$ as a function of the complex parameter $\sigma$. For each fixed $(k,\sigma)$, \eqref{eq:equation_laplace_transformed} is a boundary-value problem in $z$. When this problem admits a regular solution in the entire domain $0 \leq z \leq 1$, the resolvent is well defined. Conversely, values of $\sigma$ for which the boundary-value problem fails to have a regular solution correspond to singularities of the resolvent. We denote the set of such values of $\sigma$ for a given $k$ by $\Sigma_k$ and call it the spectrum of $\mathsfbi{M}_k$.

The spectrum $\Sigma_k$ is directly assessed by making explicit the resolvent of $\mathsfbi{M}_k$, or equivalently, solving the boundary-value problem \eqref{eq:equation_laplace_transformed} for each $(k,\sigma)$. In practice, this program reduces to solving a second-order differential equation. This point is made clear by reorganizing \eqref{eq:equation_laplace_transformed} in terms of the streamfunction as
\begin{align} \label{eq:forced_TG_equation}
\mathcal{T}_{k,\sigma} \tilde{\psi} = F \quad \text{with} \quad \mathcal{T}_{k,\sigma} \equiv \frac{d^2}{d z^2} + \frac{k^2 N^2}{(k U - \sigma)^2} - \frac{k U_{zz}}{k U - \sigma} - k^2 ,
\end{align}
where $F(z)$ collects the inhomogeneous terms involving $\hat{\bm{v}}_0$ and $\hat{\bm{f}}_n$. The homogeneous counterpart of this equation, $\mathcal{T}_{k,\sigma}\tilde{\psi}=0$, is the Taylor--Goldstein equation governing unforced internal waves in shear flows. For the inhomogeneous problem \eqref{eq:forced_TG_equation}, the solution can be written as an integral representation in terms of the corresponding Green function \citep[e.g.,][]{bender2013advanced}. We summarize this Green-function solution in Appendix~\ref{sec:app_spec}; for our purposes, the key point is that it becomes singular for particular values of $\sigma$, and these values constitute $\Sigma_k$.

As elucidated in Appendix~\ref{sec:app_spec}, the spectrum $\Sigma_k$ consists of two parts with distinct physical origins. First, isolated values of $\sigma \in \Sigma_k$ correspond to regular eigenvalue solutions of the Taylor--Goldstein equation. We collect such values of $\sigma$ to define a subset of the spectrum $\Sigma^d_k = \left\{\sigma \,\vert\, D(k,\sigma)=0\right\}$, where $D(k,\sigma)$ specifies the dispersion relation of vertically standing internal waves and is obtained by enforcing the boundary conditions on the Taylor--Goldstein equation (see Appendix~\ref{sec:app_dispersion}). Since these solutions are regular throughout the domain, the Miles--Howard stability theorem applies and ensures that $\Sigma^d_k \subset \mathbb{R}$ under the assumption $N^2/U_z^2 > 1/4$ for all $0 \leq z \leq 1$. 

Second, when $\sigma/k$ lies within the range of the background flow $U(z)$, the Taylor--Goldstein equation admits singular solutions associated with a critical level at which $kU=\sigma$ (Appendix~\ref{sec:app_continuous}). The corresponding values of $\sigma$ then form a continuous interval, $\Sigma^c_k = \left\{\sigma \,\vert\, U(0) \leq \sigma/k \leq U(1)\right\}$. In the following, we refer to $\Sigma^d_k$ and $\Sigma^c_k$ as the discrete and continuous spectra, respectively.
}

\begin{figure}
\centerline{\includegraphics[width=0.5\columnwidth]{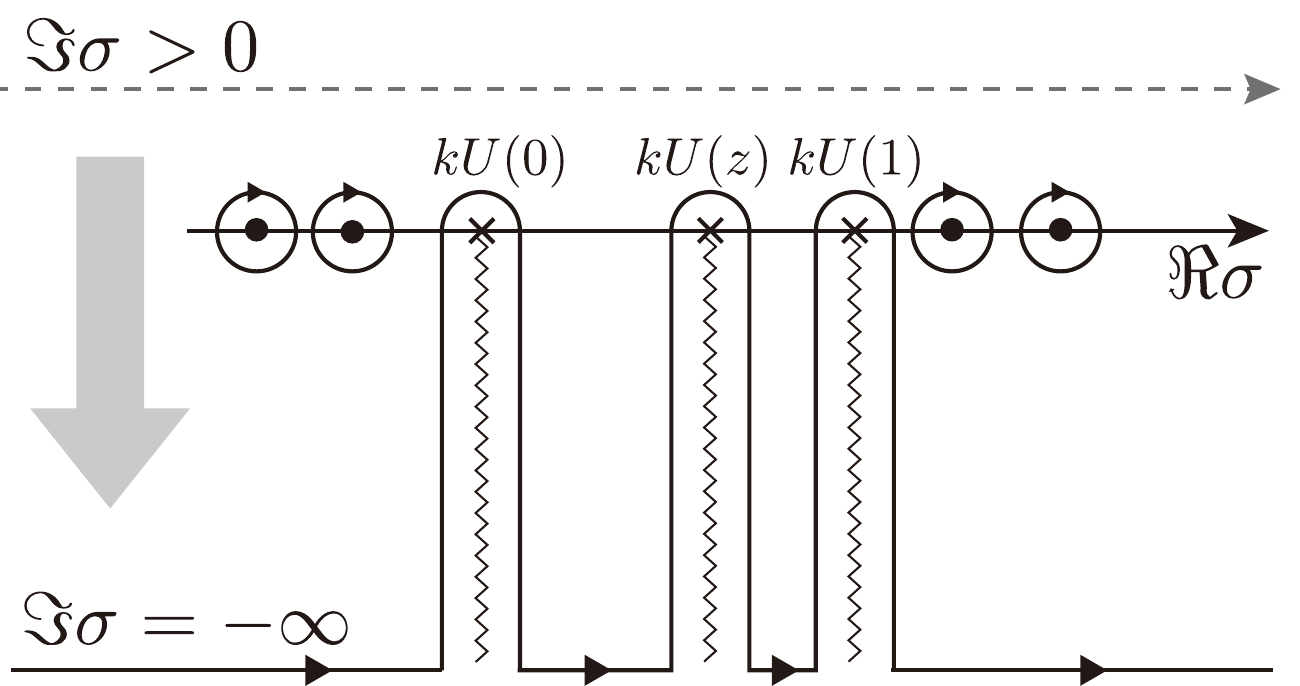}}
\caption{Integration contour of the inverse Laplace transform for a positive $k$. We shift the contour originally defined in the upper half of the complex plane in the negative direction along the imaginary axis. There exist an infinite number of poles, which involve the eigenvalues of $\mathsfbi{M}_k$, i.e., the discrete spectrum $\Sigma^d_k$, and the forcing frequency $\omega_n$. An element of the discrete spectrum, \yo{$\sigma \in \Sigma^d_k$}, is either in a range $\sigma < k U(0)$ or $\sigma > k U(1)$, while the forcing frequency $\omega_n$ may be located anywhere on the real axis. There also exist three branch points on the real axis. Two branch points $kU(0)$ and $kU(1)$ originate from the factors $\gamma^\pm(0)$ and $\gamma^\pm(1)$ and the intermediate point $kU(z)$ from $\gamma^\pm(z)$, which are involved in the solutions of the Taylor--Goldstein equation, \eqref{eq:A01}.  Since the location of the intermediate point depends on $z$ and thus ranges from $k U(0)$ to $k U(1)$, the continuous spectrum, $\Sigma^c_k$, is composed of $\sigma \in \left[ k U(0), kU(1) \right]$. }
\label{fig:integration_sigma}
\end{figure}

\yo{To examine how each element in the spectrum $\Sigma_k$ affects the solution \eqref{eq:general_solution}, we consider integration contour illustrated in Fig.~\ref{fig:integration_sigma}; we shift the contour in the negative direction of the imaginary axis but circumventing the singular points (elements of $\Sigma_k$).} The net contribution to the integration is constituted of the residues from poles and integration along branch cuts. In the first term in \eqref{eq:general_solution}, which represents the free wave solution, the residues of $\sigma \in \Sigma^d_k$ represent standing waves that are the discrete eigenmodes of the system, in agreement with the fact that $D(k, \omega) = 0$ represents the dispersion relations of these modes. On the other hand, integration along a branch cut that extends from a point in $\Sigma^c_k$ originates from the steady shear flow. There exist three branch points as indicated in Fig.~\ref{fig:integration_sigma}, among which the middle one dependent on $z$ reflects the critical level. This point appears in the integrand in the form
\begin{align}
\left[ \sigma - k U(z) \right]^{m + 1/2 \pm \ii \nu} ,
\end{align}
where $m$ is an integer, and $\nu$ is a function of $\sigma$ and $k$ (see Appendix~\ref{sec:app_spec}). Because the exponent is not an integer, the residue theorem is not applicable there. An asymptotic evaluation shows that the contributions from the branch cuts behave algebraically in time, and the wave energy will not grow from a smooth initial profile in the long-term limit \citep{brown1980algebraic,jose2015analytical}. The unforced system is, therefore, stable.

The integrand of the second term in \eqref{eq:general_solution} possesses another singularity at the forcing frequency, $\sigma = \omega_n$. If $\omega_n$ does not belong to the spectrum $\Sigma_k$, the residue of this pole yields harmonic oscillation, and hence the system is still stable. However, if the forcing frequency matches an element of the spectrum, integration around this point can exhibit algebraic growth in time. To focus on this resonant process, it is convenient to add a small relaxation term that mimics viscosity to the time derivative, setting $\partial_t \to \partial_t + r$ with $r > 0$ in \eqref{eq:equation_Mk}. In this system, $\Sigma_k$ is slightly lowered along the imaginary axis so that the contributions to the integral in \eqref{eq:general_solution} from the spectrum are exponentially damped with time. The residue from $\sigma = \omega_n$ is solely responsible for the long-term behavior. Accordingly, the stationary response solution is derived as
\begin{align} \label{eq:stationary}
\hat{\bm{v}} = \sum_{n = - \infty}^\infty \frac{e^{- \ii \omega_n t} \hat{\bm{f}}_n}{ (\omega_n + \ii r) \mathsfbi{I} - \mathsfbi{M}_k} .
\end{align}
This solution is regular everywhere for a finite $r$, but when taking the inviscid limit $r \to 0$, singularity can arise.
Specifically, if the forcing frequency $\omega_n$ belongs to the discrete spectrum $\Sigma^d_k$, the solution diverges almost everywhere in $z$. On the other hand, if $\omega_n$ is an interior point of the continuous spectrum $\Sigma^c_k$, the solution is singular at a particular level (Appendix~\ref{sec:app_continuous}). In either case, wave energy $E$ \yo{diverges}---energy is unboundedly supplied from the forcing term.
% \footnote{For a Rossby wave case, the wave energy does not diverge at a critical layer, but the pseudoenergy does. The present consideration qualitatively applies to various shear flow problems.}
The spectrum inherent in the unforced system determines the condition of the permanent energy supply from the external tidal force.

Although we are mainly interested in the inviscid problem, keeping a small but finite $r > 0$ through the paper is useful in that it moves a location of a singular point slightly from a real axis in the $z$- or $k$-plane. This enables us to (i) display the vertical structure of a solution that possesses a critical level (Fig.~\ref{fig:vertical_structure}d), (ii) determine which half-plane a pole or branch point exists in (Fig.~\ref{fig:integration_k}), (iii) learn the behavior of a solution in the genuine resonant situation (Fig.~\ref{fig:spectrum_cg0}), and (iv) derive the explicit formula for the energy production rate \yo{(Section~\ref{sec:energy_production})}.

\section{\yo{Stationary solution from a boundary value problem}} \label{sec:stationary_solution}

\yo{
The stationary solution in the form \eqref{eq:stationary} is informative for a direct qualitative physical interpretation of the linear response to external forcing. However, \eqref{eq:stationary} is not convenient for explicit calculations because the resolvent operator involves an intricate integral transform \eqref{eq:phi_green}. We therefore proceed by solving an equivalent one-dimensional boundary-value problem: combining the potential and vortical parts yields a single Taylor--Goldstein equation with an inhomogeneous bottom boundary condition, which we solve for each tidal harmonic.

Having obtained these stationary solutions, we interpret and classify them using the spectrum identified in Section~\ref{sec:spectrum}. In particular, whether the forcing frequency intersects the discrete spectrum $\Sigma^d_k$ or lies within the continuous spectrum $\Sigma^c_k$ determines the qualitative character of the response (standing modes versus critical-level behaviour). For spatially localized topography, this spectral classification becomes particularly transparent in the far field: an asymptotic evaluation of the inverse Fourier integral separates pole contributions associated with the discrete spectrum from branch-cut contributions associated with the continuous spectrum (Section~\ref{sec:localized_topography}). This viewpoint provides a direct link between the initial-value resolvent formulation of Section~\ref{sec:general_solution} and the boundary-value approach developed below.
}

\subsection{\yo{Single wavenumber solution}}
Now, we inherit the idea of the previous section that the model involves a small relaxation term specified by $r > 0$, and the system reaches a stationary state. Consequently, the solution in the inviscid limit consists of the linear superposition of harmonic oscillations as
\begin{align} \label{eq:solution_all}
\left[ \hat{\psi}_1(k, z, t), \hat{\rho}_1(k, z, t) \right] = \lim_{r \to 0} \sum_{n = -\infty}^\infty \left[\hat{\psi}_{1n} (z; k, r), \hat{\rho}_{1n} (z; k, r) \right] e^{- \ii \omega_n t}.
\end{align}
Each element of the streamfunction solves the homogeneous Taylor--Goldstein equation, $\mathcal{T}_{k,\omega_n + \ii r} \hat{\psi}_{1n} = 0$, with the inhomogeneous bottom boundary condition \eqref{eq:boundary_Bessel} and the homogeneous upper boundary condition. The density perturbation is then related via \eqref{eq:equation_rho_k}. We readily write down the solutions as
\begin{subequations} \label{eq:solution_single}
\begin{align}
\hat{\psi}_{1n} (z; k, r) & = \frac{(k U_0 - \omega_n) \hat{h}(k)}{k} J_n \left( \frac{k U_T}{\omega_T} \right) \frac{A_1(z; k, \omega_n + \ii r)}{D(k, \omega_n + \ii r)} \\
\hat \rho_{1 n}(z; k, r) & =\frac{k N^2}{kU(z)-(\omega_n+ \ii r)} \hat \psi_{1 n} (z; k, r) ,
\end{align}
\end{subequations}
where $A_1$ represents a vertical structure function that is a solution for $\mathcal{T}_{k,\omega_n + \ii r} A_1 = 0$ with a one-side boundary condition $\left. A_1 \right\vert_{z=1} = 0$, and $D$ plays the role of normalization such that $\left. A_1 \right\vert_{z=0} = D$ is understood. Since the solutions in physical space, $\psi$ and $\rho$, are real functions, their Fourier transform should satisfy the symmetric property as to changes in signs of wavenumber and frequency, i.e., $\hat{\psi}_{1n} (z; k, r) = \hat{\psi}^\dag_{1-n} (z; - k, r)$ and $\hat{\rho}_{1n} (z; k, r) = \hat{\rho}^\dag_{1-n} (z; - k, r)$, where ${}^\dag$ denotes the complex conjugate. Therefore, it is enough to consider the cases of $n \geq 0$ in the following.

\begin{figure}[t]
\centering\includegraphics[bb=0 0 552 444, width=\columnwidth]{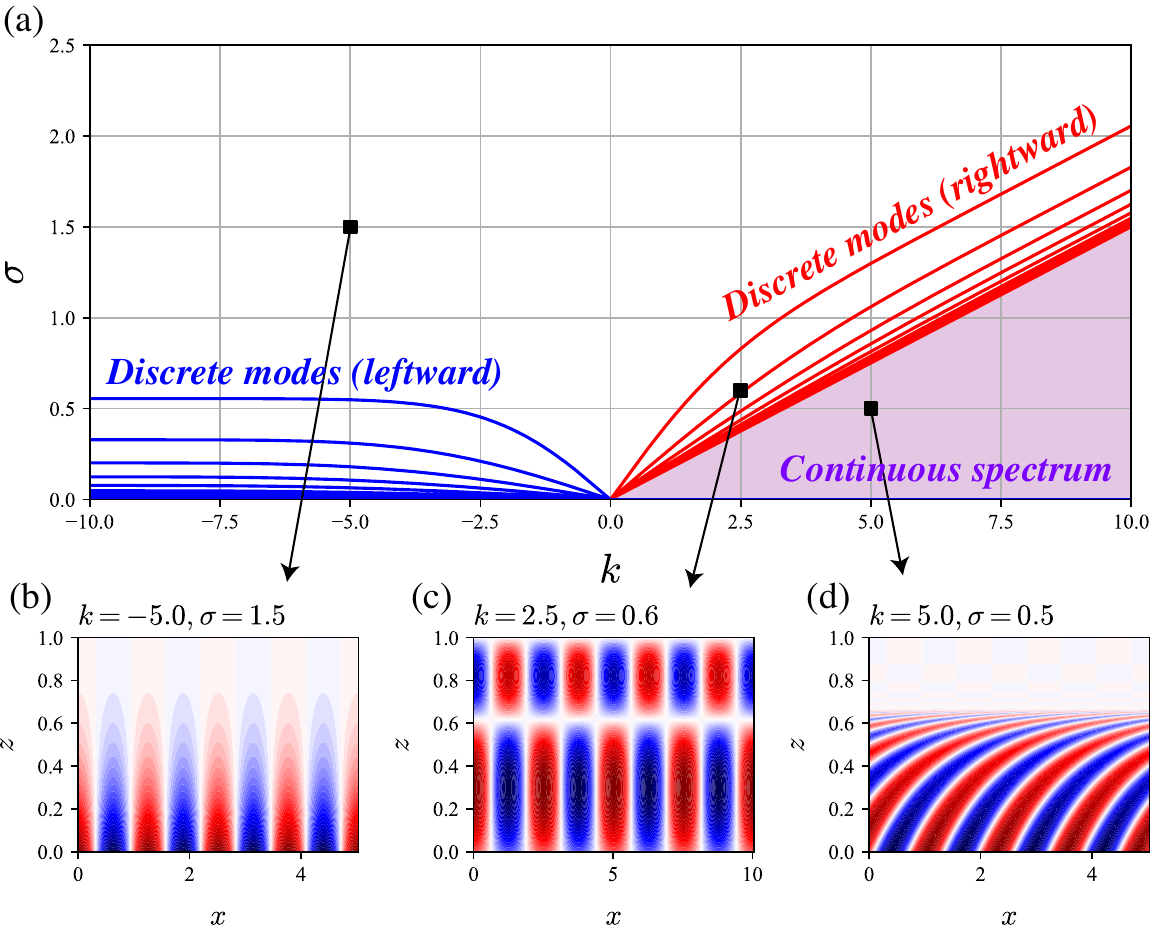}
\caption{\label{fig:vertical_structure} (a) The spectrum of an operator $\mathsfbi{M}_k$ for a constant shear and stratification case, $N = 1$ and $U = 0.15 z$. \yo{Blue and red curves correspond to $D(k, \sigma) = 0$ and represent the dispersion relations of vertically standing eigenmodes. The magenta region represents the continuous spectrum.} (b-d) Spatial structures of stationary solutions for \yo{each horizontal wavenumber} are demonstrated by drawing the real part of $A_1(z; k, \sigma) e^{\ii k x}$.}
\end{figure}

Equation $\mathcal{T}_{k,\sigma} A_1 = 0$ can be solved for a general set of $N(z)$ and $U(z)$ numerically, but for the simplest case of uniform shear and stratification, $N=1$ and $U = U_0 + (U_1 - U_0) z$, an analytical expression is available (Appendix~\ref{sec:app_uniform_shear}). Figure~\ref{fig:vertical_structure} shows \yo{the spectrum of $\mathsfbi{M}_k$ as well as some examples of the vertical structure function $A_1 (z)$} where we set $U_0 = 0$ and $U_1 = 0.15$. \yo{In these plots, we augment the structure function by multiplying $e^{\ii k x}$ to illustrate both the vertical and horizontal structures.} Depending on the horizontal wavenumber and the forcing frequency, the wave structure exhibits distinct characteristics. For $(k, \sigma) = (-5, 1.5)$ as in Fig.~\ref{fig:vertical_structure}b, because the generated wave's frequency is greater than the background buoyancy frequency, the solution exhibits an evanescent structure localized close to the bottom. If we choose a wavenumber and frequency close to a dispersion curve of a discrete mode \yo{(i.e., the discrete spectrum $\Sigma^d_k$)}, specifically $(k, \sigma) = (2.5, 0.6)$ in Fig.~\ref{fig:vertical_structure}c, the solution becomes a vertically standing wave structure\footnote{\yo{We cannot define a stationary solution if the forcing frequency lies in the dispersion curves since a denominator $D$ in \eqref{eq:solution_single} vanishes in that limit.}}. There exist an infinite number of such standing modes in the vicinity of the edges of the continuous spectrum \yo{(Appendix~\ref{sec:app_accumulation})}.

\yo{By contrast,} when the forcing frequency \yo{lies within the continuous spectrum}, \yo{$\omega_n \in \Sigma^c_k$}, the \yo{inviscid stationary} solution \yo{exhibits a singularity} at a critical level where $k U$ coincides with $\omega_n$. More specifically, according to \eqref{eq:A01} and \eqref{eq:gamma_critical}, the vertical structure function $A_1$ \yo{admits a local representation}
\begin{align} \label{eq:A_branch}
A_1 (z; k, \omega_n + \ii r) & = (z - z^c)^{1/2 - \ii \nu} \alpha_1(z) + (z - z^c)^{1/2 + \ii \nu} \alpha_2(z) ,
\end{align}
where $z^c$ and $\nu$ are functions of $k$ and $\sigma = \omega_n + \ii r$ defined through
\begin{align}
k U(z^c) = \sigma \quad \text{and} \quad \nu = \sqrt{\left. \frac{N^2}{U_z^2} \right\vert_{z = z^c} - \frac{1}{4}} ,    
\end{align}
and $\alpha_1(z)$ and $\alpha_2(z)$ are smooth functions. In the inviscid limit $r \to 0$, the branch point $z = z^c$ \yo{approaches the real axis and the two local solutions in \eqref{eq:A_branch} are connected across $z^c$ according to the rule \eqref{eq:connection_rule}. The upper-boundary condition then imposes $\alpha_2 = \alpha_1^\dag$.} Above the critical level, since the two terms in \eqref{eq:A_branch} have the same amplitude, $A_1$ exhibits a standing wave structure. Below the critical level, the amplitudes of the two terms differ by a factor $e^{2 \pi \nu}$, so that the upward-propagating \yo{component} dominates the downward-propagating \yo{component}. The wave amplitude in the upper layer is smaller than that of the lower layer by a factor $e^{\pi \nu}$. These features are evident in Fig.~\ref{fig:vertical_structure}d, $(k, \sigma) = (5, 0.5)$.

\yo{
This amplitude jump across the critical level is a classical signature of wave focusing and absorption into the background flow at a critical layer, most transparently discussed in initial-value problems \citep{booker1967critical}. The discussion above, however, assumes a single horizontal wavenumber, which corresponds to a sinusoidal topography. For localized topography, the forcing excites a continuum of horizontal wavenumbers, and the manifestation of the critical-layer behaviour is qualitatively different, as we discuss below.
}

\subsection{Wave generation from localized topography} \label{sec:localized_topography}

\begin{figure}[t]
\centering\includegraphics[width=0.5\columnwidth]{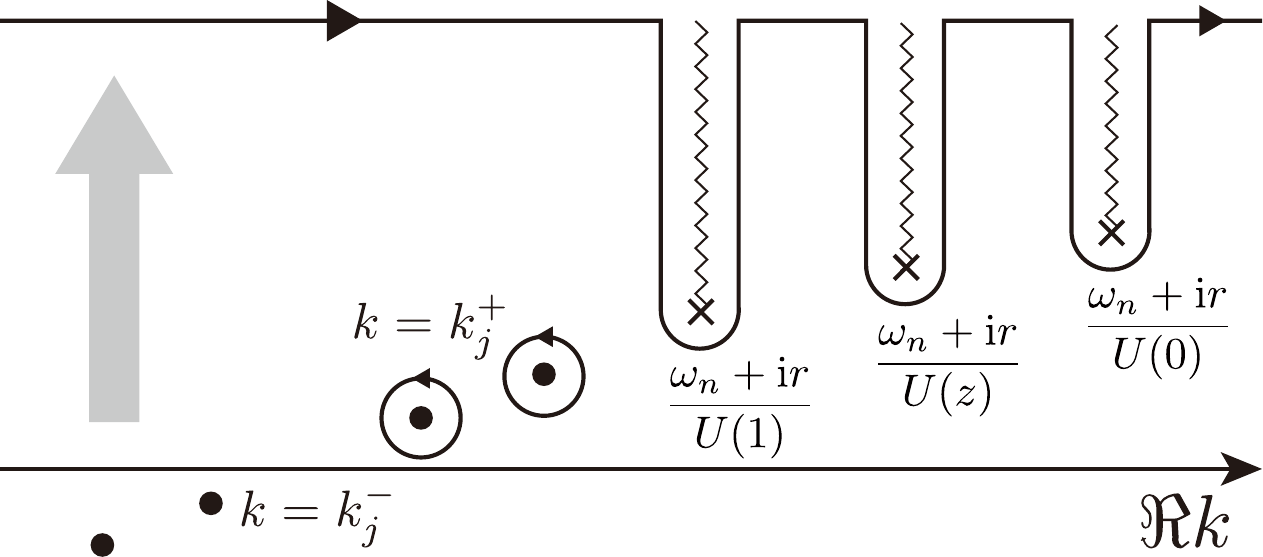}
\caption{\label{fig:integration_k} Integration contour of the Fourier integral \eqref{eq:Fourier_transform} for positive $x$. The contour originally defined on the real axis is shifted over a finite distance on the complex plane along the imaginary axis. For a wavenumber $k = k_r + \ii k_i$, the exponential factor in the integrand becomes $e^{\ii k x} = e^{\ii k_r x - k_i x}$. Therefore, for a large $x$, contributions from the line segments in the upper half of the complex plane become exponentially small, and the integration is dominated by those around the poles and branch points, which approach the real axis in the inviscid limit $r \to 0$. For negative $x$, the integration contour is shifted in the opposite direction. Note that the branch point around which the solution \eqref{eq:algebraic_solutions} is evaluated is $k' = (\omega_n + \textrm{i} r) / U(z)$. Currently, we are considering the case when $U>0$ for any $z \in [0, 1]$. For $U < 0$, the corresponding branch point exists in the lower half-plane.
}
\end{figure}

\yo{
We now turn to wave generation by bottom topography that is localized in physical space. In this case, the full solution is obtained by performing the inverse Fourier transform, i.e., inserting \eqref{eq:solution_single} and \eqref{eq:solution_all} into \eqref{eq:Fourier_transform}. This step requires some care compared with the inverse Laplace transform discussed earlier. Even when the integrand is analytic in $k$ along the real axis, shifting the contour arbitrarily far into the complex plane is generally inappropriate because the analytic structure of $\hat{h}$ must also be taken into account. For instance, for a Gaussian ridge $h(x) \propto e^{-x^2/2}$, the Fourier transform behaves as $\hat{h}(k) \propto e^{-k^2/2}$ and diverges as $k \to \pm \ii \infty$. Closed-form expressions are therefore available only for special choices of $U$, $N$, and $h$. Here, we instead adopt an asymptotic approach to describe the solution far from the localized topography. As illustrated in Fig.~\ref{fig:integration_k}, the integration contour is shifted a finite distance into the complex plane while circumventing the relevant singularities. In the limit $|x| \to \infty$, the dominant contributions arise from singularities located near the real axis, since the integrand decays exponentially away from it. This contour-deformation argument applies to a broad class of spatially localized topographies.
}

\yo{In the present problem, the relevant singularities are poles and branch points, which originate from the discrete and continuous spectra of the system, respectively.} We first investigate the contributions from the poles. For a prescribed real $k$, the location of the zeros of $D(k, \sigma)$ is specified by the discrete spectrum as $\Sigma^d_k = \{ \sigma^\pm_m \}_{m = 1, 2, \ldots}$, where $\sigma^+$ and $\sigma^-$ are distinguished in terms of the sign of $\sigma / k - U$ \yo{(Appendix~\ref{sec:app_spec_classification})}. Similarly, for a prescribed real $\sigma$, we shall write the location of the zeros of $D(k, \sigma)$ on the real axis as $K^d_\sigma = \{ k_j \}_{j = 1, 2, \ldots}$. When $r > 0$ is sufficiently small, in the vicinity of $k = k_j \in K^d_{\omega_n}$, $D$ is locally expanded as
\begin{subequations}
\begin{align}
D(k, \omega_n + \ii r) & = - \left[ k - k_j - \frac{\ii r}{c^g_j} + \mathcal{R}(k, r) \right] c^g_j D_\sigma \label{eq:D_expand} \\
\mbox{with} \quad c^g_j & \equiv - \frac{D_k}{D_\sigma} = \left. \frac{\partial \sigma}{\partial k} \right\vert_{D=0} \label{eq:group_velocity} ,
\end{align}
\end{subequations}
where $\mathcal{R}$ is a residual factor that vanishes in the limits of $r \to 0$ and $k \to k_j$ faster than the other terms, i.e., $\mathcal{R} = o(k - k_j, r)$ in Landau notation. Expression \eqref{eq:group_velocity} represents the slope of the dispersion curve, i.e., the horizontal group velocity of an eigenmode. According to \eqref{eq:D_expand}, the integrand of the Fourier integral \eqref{eq:Fourier_transform} involves a pole at $k = k_j + \ii r / c_j^g + o(r) \equiv k'_j$. If $c_j^g > 0$, this pole is located slightly above the real axis, and the circular integration around it contributes to the solution for $x \to + \infty$. On the other hand, if $c_j^g < 0$, the pole is located below the real axis, and it contributes when $x \to - \infty$. We shall write these factors involved in the Fourier integral in \eqref{eq:Fourier_transform} as
\begin{align} \label{eq:standing_wave}
\lim_{r \to 0} \frac{1}{2 \pi} \oint_{k \sim k'_j} \hat{\psi}_{1n} e^{\ii k x} dk = \frac{(k_j U_0 - \omega_n) \hat{h}(k_j)}{\ii k_j \vert c^g_j \vert D_\sigma(k_j, \omega_n)} J_n \left( \frac{k_j U_T}{\omega_T} \right) A_1(z; k_j, \omega_n) e^{\ii k_j x} ,
\end{align}
where integration is performed in the counterclockwise (clockwise) direction for $c^g_j > 0$ ($c^g_j < 0$). Equation \eqref{eq:standing_wave} represents a vertically standing-mode wave train that travels far away from the \yo{topography} while not changing its amplitude. This is the solution discussed by \cite{lamb2018internal}.

\begin{figure}
\centerline{\includegraphics[width=0.8\columnwidth]{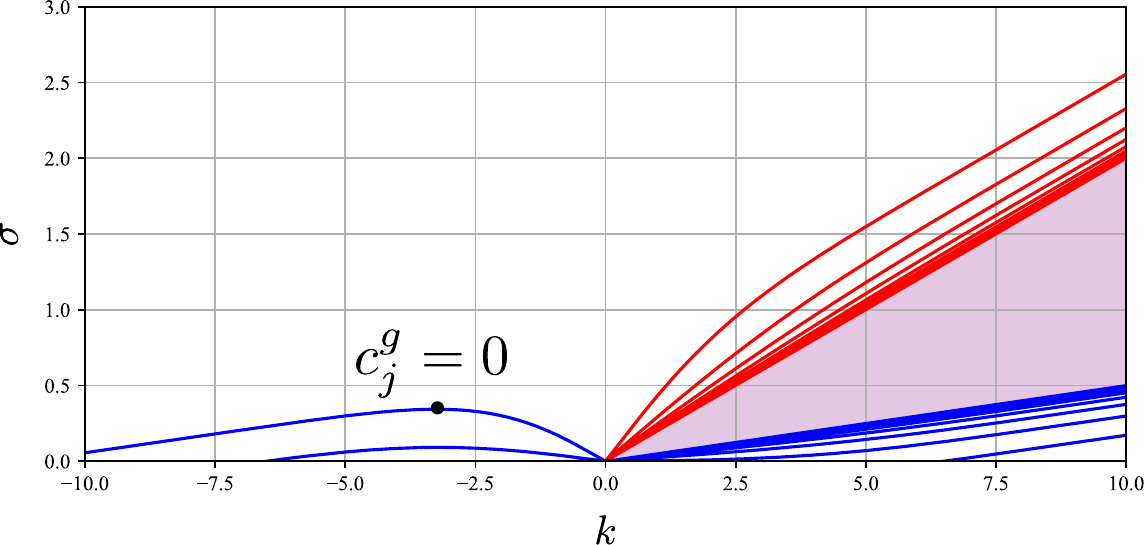}}
\caption{The spectrum of an operator $\mathsfbi{M}_k$ for a constant shear and stratification case, $N = 1$ and $U = 0.05 + 0.15 z$. On the dispersion curves of discrete modes, there exist points where group velocity $c^g_j$ vanishes.
}
\label{fig:spectrum_cg0}
\end{figure}

Equations \eqref{eq:D_expand} and \eqref{eq:standing_wave} are invalid when the group velocity vanishes, $c^g_j = 0$. An example of this situation is illustrated in Fig.~\ref{fig:spectrum_cg0}. An alternative expansion of $D$ in this case is
\begin{align} \label{eq:D_expand_2}
D(k, \omega_n + \ii r) = - \left[ \frac{(k - k_j)^2}{2} - \frac{\ii r}{c^g_{j k}} + \mathcal{R}(k, r) \right] c^g_{j k}  D_\sigma ,
\end{align}
where $\mathcal{R} = o((k - k_j)^2, r)$, and $c^g_{j k} = - D_{kk} / D_\sigma$ is understood. Supposing $c^g_{jk} < 0$, the integrand in \eqref{eq:D_expand_2} has a pair of poles at
\begin{align} \label{eq:pole_when_c0}
k = k_j \pm (1 - \ii) \sqrt{\frac{r}{\vert c^g_{jk} \vert}} + o\left( r^{1/2} \right) \equiv k^\pm_j .    
\end{align}
Circular integration around these poles for small $r$ is evaluated as
\begin{align}
\frac{1}{2 \pi} \oint_{k \sim k^\pm_j} \hat{\psi}_{1n} e^{\ii k x} dk & = \frac{(k_j U_0 - \omega_n) \hat{h}(k_j)}{(1 + \ii) k_j  \sqrt{r \vert c^g_{j k} \vert} D_\sigma(k_j, \omega_n)} J_n \left( \frac{k_j U_T}{\omega_T} \right) A_1(z; k_j, \omega_n) e^{\ii k^\pm_j x} \nonumber \\
& + o \left( r^{-1/2} \right) .
\end{align}
Accordingly, the stationary solution is divergent in the inviscid limit with the power law $\vert \psi \vert \propto r^{-1 /2}$. Physically, if the group velocity is 0, wave energy accumulates locally above the \yo{topography}. Interference between the accumulating wave and the topographic forcing results in resonant amplification\footnote{\yo{This situation does not correspond to a convective or absolute instability, as no exponentially growing modes exist when $Ri>1/4$ (Miles--Howard theorem). Instead, the vanishing group velocity indicates a resonance condition, where the forced response is amplified due to the matching of the forcing frequency with the natural frequency of a stationary wave mode.}}. Far from the \yo{topography}, the wave amplitude is damped exponentially over the length scale $\sqrt{\vert c^g_{jk} \vert / r}$ according to \eqref{eq:pole_when_c0}. In this process, the group velocity dispersion represented by the curvature of the dispersion curve, $c^g_{jk}$, plays distinctive roles. When $\vert c^g_{jk} \vert$ is small, a broad range of wavenumbers are involved in the resonance so that the wave is much more amplified. At the same time, the group velocity of these side-band components is still small, which results in shortening the decay length scale.

We next consider the contributions from the branch cut. For a prescribed $k$ and $\sigma$, $A_1(z; k, \sigma)$ can have a branch point at $z = z'$ where $k U(z') = \sigma$. In the same way, for a prescribed $\sigma$ and $z$, a branch point $k = k'$ is defined where $k' U(z) = \sigma$ is satisfied. At a particular level where $U(z) = 0$, however, this consideration does not make sense because $k U(z) = \sigma$ is identically satisfied for any $k$ if $\sigma = 0$. This location is the critical level of the steady component, $n = 0$, where an inviscid stationary solution is undefinable. In the following, we neglect this level and set $U(z) \neq 0$ and $n > 0$.

Now, we write the branch point corresponding to the critical level as $k' (z, r) = (\omega_n + \ii r) / U(z)$. Around this point, inserting $z = z^c (k', \omega_n + \ii r)$ and $z^c = z^c (k, \omega_n + \ii r)$ into \eqref{eq:A_branch}, we expand $A_1$ as
\begin{align*}
A_1(z; k, \omega_n + \ii r) & = \left( \frac{U}{k' U_z} \right)^{1/2 - \ii \nu} \left(k - k' \right)^{1/2 - \ii \nu} \alpha_1(z) \\
& + \left( \frac{U}{k' U_z} \right)^{1/2 + \ii \nu} \left(k - k' \right)^{1/2 + \ii \nu} \alpha_2(z) + \mathcal{R} ,
\end{align*}
where $\partial_k z^c = - U(z^c) / [k U_z(z^c)]$ has been used. Note that $\nu$ was a function of $z^c(k, \omega_n + \ii r)$ in \eqref{eq:A_branch}, but it is now a function of $z$. This replacement yields an error $\propto (k - k')^{3/2 \pm \ii \nu} \log (k - k')$, which is negligible in the following result and hence included in the residual term $\mathcal{R}$.
For $U(z) > 0$, the branch point is located in the upper half of the complex $k$ plane. Its contribution to the integral at $x \gg 1$ is evaluated as
\begin{subequations} \label{eq:algebraic_solutions}
\begin{align}
\lim_{r \to 0} \frac{1}{2 \pi} \int_{k \sim k'} \hat{\psi}_{1n} e^{\ii kx} dk & = \left[ \left( \frac{U^2}{\omega_n U_z} \right)^{1/2 - \ii \nu} \frac{e^{- \pi \nu / 2} \alpha_1 x^{\ii \nu}}{\Gamma(- 1/2 + \ii \nu)} + \left( \frac{U^2}{\omega_n U_z} \right)^{1/2 + \ii \nu} \frac{e^{\pi \nu / 2} \alpha^\dag_1 x^{- \ii \nu}}{\Gamma(- 1/2 - \ii \nu)} \right] \nonumber \\
\times & (U_0 - U) \hat{h} \left( \frac{\omega_n}{U} \right) J_n \left( \frac{n U_T}{U} \right) \frac{\ii^{- 1/2} x^{- 3/2} e^{\ii \omega_n x / U}}{D(\omega_n / U, \omega_n)}  + o\left( x^{-3/2} \right) \label{eq:algebraic_psi} \\
\lim_{r \to 0} \frac{1}{2 \pi} \int_{k \sim k'} \hat{\rho}_{1n} e^{\ii kx} dk & = \left[ \left( \frac{U^2}{\omega_n U_z} \right)^{1/2 - \ii \nu} \frac{e^{- \pi \nu / 2} \alpha_1 x^{\ii \nu}}{\Gamma(1/2 + \ii \nu)} + \left( \frac{U^2}{\omega_n U_z} \right)^{1/2 + \ii \nu} \frac{e^{\pi \nu / 2} \alpha_1^\dag x^{- \ii \nu}}{\Gamma(1/2 - \ii \nu)} \right] \nonumber \\
\times & \frac{ (U_0 - U) \omega_n N^2}{U^2} \hat{h} \left( \frac{\omega_n}{U} \right) J_n \left( \frac{n U_T}{U} \right) \frac{\ii^{1/2} x^{- 1/2} e^{\ii \omega_n x / U}}{D(\omega_n / U, \omega_n)}  + o\left( x^{-1/2} \right) ,
\end{align}
\end{subequations}
where an asymptotic formula in Appendix~\ref{sec:app_asymptotic_formula} is used. Note that the exponents on $x$ in \eqref{eq:algebraic_solutions} match those reported by \cite{camassa2013transient}, who dealt with time-periodic but unforced disturbances in a stratified shear flow. The present study demonstrates that a forced problem solved with a different asymptotic approach yields consistent results.

\begin{figure}[t]
\centering\includegraphics[width=0.8\columnwidth]{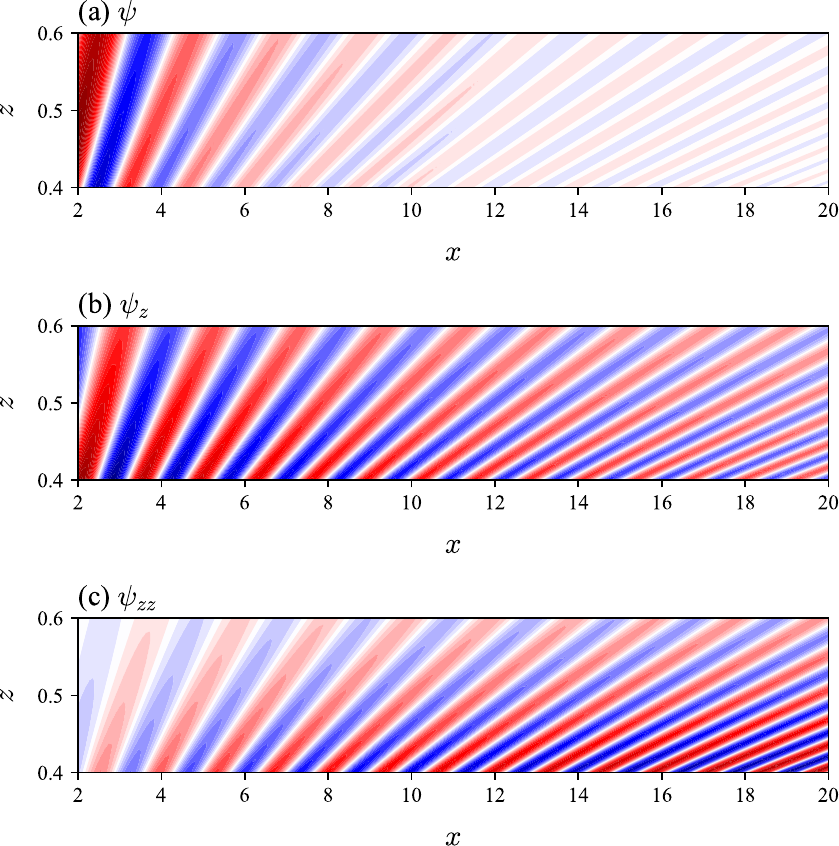}
\caption{\label{fig:algebraic} Illustration of the asymptotic formula \eqref{eq:algebraic_psi}. Snapshots of $\psi_1$, $\partial_z \psi_1$, and $\partial_z^2 \psi_1$ are computed for $N=1$, $U=0.5z$, $n=1$, $\omega_T=1$, $U_T = 0.1$, $\hat{h}=1$ (i.e., $h = \delta(x)$), and $\alpha$ presented as \eqref{eq:alpha_Bessel}.}
\end{figure}

A marked difference of \eqref{eq:algebraic_solutions} from the standing-mode wave solutions \eqref{eq:standing_wave} is found in their spatial structure. The present solutions exhibit algebraic dependence on $x$. Furthermore, Owing to the factor $U(z)$, the solution is not separable into the $x$- and $z$-dependent parts. As a result, $\psi$, $\psi_z$, and $\psi_{zz}$ obey different power laws. The perturbations of velocity, density, and vertical shear exhibit the dependence on $x \gg 1$ roughly as 
\begin{align} \label{eq:algebraic_growth}
\vert w_1 \vert \propto x^{- 3/2} , \quad \vert u_1 \vert, \vert \rho_1 \vert \propto x^{- 1/2}, \quad \left\vert \frac{\partial u_1}{\partial z} \right\vert \propto x^{1/2} .
\end{align}
While velocity and density decay, the velocity gradient grows far from the \yo{topography}. Figure~\ref{fig:algebraic} illustrates the asymptotic formula \eqref{eq:algebraic_psi}. Notably, the structure of the solution differs from that in the single-wavenumber case. Now, the equiphase curves are almost straight, and there does not exist a level where the solution is discontinuous. The angle of the equiphase lines varies in the horizontal direction, reflecting the increase of the vertical wavenumber and, consequently, the growth of the velocity gradient. 

\yo{The algebraic dependence of the asymptotic solutions in \eqref{eq:algebraic_growth} can be interpreted in terms of conservation of wave activity (pseudomomentum), a quadratic invariant associated with horizontal translation symmetry.
A formal definition and the corresponding conservation law are given later in Section~\ref{sec:pseudomomentum}; here we use only the qualitative implication that, once the disturbance has left the forcing region, the pseudomomentum carried by the wave field is conserved during propagation.
In the far field, pseudomomentum is dominated by the product of the density perturbation and the vorticity, $\rho_1 \nabla^2 \psi_1/N^2$ (up to subdominant correction terms), so if $\rho_1$ decays it must be compensated by the growth of $\nabla^2 \psi_1$.
Because $w_1$ decays much faster than $u_1$, the term $w_{1x}$ in $\nabla^2 \psi_1=-u_{1z}+w_{1x}$ is subdominant, and hence $\nabla^2 \psi_1 \approx -u_{1z}$.
Consequently, pseudomomentum conservation suggests $\rho_1 u_{1z}=\mathcal{O}(1)$ as $x\to\infty$, consistent with \eqref{eq:algebraic_growth}.

By contrast, wave energy $E$ and enstrophy $(\nabla^2 \psi_1)^2$ are not conserved in a stratified shear flow. Indeed, \eqref{eq:algebraic_growth} tells that the energy decreases while the enstrophy increases with $x$, respectively. Here, a difference from a single wavenumber solution is worth noting; for a singular solution \eqref{eq:A_branch}, the power laws around the critical level are $\vert w_1 \vert \propto \vert z - z^c \vert^{1/2}$, $\vert u_1 \vert, \vert \rho_1 \vert \propto \vert z - z^c \vert^{-1/2}$, and $\vert u_{1z} \vert \propto \vert z - z^c \vert^{-3/2}$. Accordingly, the wave energy density algebraically grows towards the critical level. On the other hand, for the present regular solution \eqref{eq:algebraic_solutions} composed of a continuous range of wavenumbers, the wave energy density decays during the horizontal propagation. In either case, because the velocity shear grows during propagation, waves will eventually break, resulting in energy dissipation.}

In addition to $k = (\omega_n + \ii r) / U(z)$, the integrand of the Fourier integral involves other branch points at $k = (\omega_n + \ii r) / U(0)$ and $k = (\omega_n + \ii r) / U(1)$. Differing from the result above, their contributions for a large $\vert x \vert$ are separable with respect to $x$ and $z$. This study does not discuss the contributions to the solutions from these points.

\section{\yo{Energy conversion rates}} \label{sec:energetics}

\yo{In the previous section, we derived stationary solutions for internal waves generated over localized, small-amplitude topography. In this setting, mechanical energy is continuously converted from the prescribed barotropic motions (a steady current and an oscillatory tide) to baroclinic disturbances. The goal of the present section is to quantify this barotropic-to-baroclinic conversion rate in a form suitable for mixing and wave-drag parameterizations.

To derive a consistent conversion-rate budget, we first note that the definition of energy---and, in particular, the partition between background and disturbance energies---is, in general, reference-frame dependent in wave--mean-flow interaction theory \citep{buhler2014waves}. Accordingly, we work in the frame moving with the oscillatory barotropic flow, as introduced in Section~\ref{sec:problem_setup}. This choice is especially convenient here because it allows the contributions of the steady current and of the tidal forcing to the conversion rate to be identified and evaluated separately. We begin by establishing the corresponding total energy budgets from the fully nonlinear equations in this co-moving frame.

In the standard no-shear setting (resting background), the energetics is straightforward: the baroclinic energy is the sum of kinetic energy and available potential energy, which is quadratic in the wave amplitude and locally conserved in the inviscid limit. The conversion rate can then be evaluated directly from the radiated internal-wave energy flux away from the topography \citep{petrelis2006tidal}.

The situation changes once a steady background flow with vertical shear is present: internal waves exchange energy and momentum with the mean flow during propagation. As a result, the wave energy alone does not form a closed budget, and defining a net barotropic-to-baroclinic conversion rate requires accounting for the mean-flow work as well as the radiated wave flux. \cite{lamb2018internal} discussed this issue in a closely related setting and proposed an energy-flux expression for the radiating discrete modes. However, an energy conversion formula applicable to singular modes associated with critical-layer processes has not yet been established, motivating the theoretical development below.

In Section~\ref{sec:total_energy_budget}, we examine the energetics of our model and derive a general conversion-rate expression that incorporates contributions from both the discrete and continuous spectra. To represent the mean-flow contribution in a compact and physically interpretable manner, in Section~\ref{sec:pseudomomentum}, we make use of the wave-activity diagnostics of pseudomomentum and pseudoenergy. Finally, in Section~\ref{sec:energy_production}, focusing on the stationary response considered in Section~\ref{sec:stationary_solution}, we obtain an explicit formula for the \emph{time-averaged} conversion rate \eqref{eq:energy_conversion_rate_n} and evaluate it numerically for a representative configuration, separating the contributions from the discrete and continuous spectra.}

\subsection{Total energy budgets} \label{sec:total_energy_budget}

\yo{We consider the fully nonlinear model \eqref{eq:equation_disturbance} applied with the nondimensionalization in Section~\ref{sec:scaling}. For this system,} the total amounts of kinetic energy and potential energy in the frame moving with the oscillating barotropic flow are represented by
\begin{subequations} \label{eq:energy_total}
\begin{align}
\mathcal{E}_K & = \iint_{\epsilon h}^1 \left[ \frac{1}{2}\left( U - \frac{\partial \psi}{\partial z} \right)^2 + \frac{1}{2} \left( \frac{\partial \psi}{\partial x} \right)^2 \right] dz dx \label{eq:total_kinetic_energy} \\
\text{and} \quad \mathcal{E}_{P} & = \iint_{\epsilon h}^1 z \rho_{tot} dz dx , \label{eq:totak_potential_energy}
\end{align}
\end{subequations}
respectively, where the lower bound of the vertical integration $\epsilon h$ depends on $x$ and $t$, and a sufficiently large extent is taken in the horizontal direction to cover the whole range in which the disturbance exists. The total density is constituted of the laminar reference part and the disturbances as $\rho_{tot} = \rho_0(z) + \rho(x,z,t)$ with $\rho_{0z} = - N^2$ understood.

We shall expand the total energy \eqref{eq:energy_total} in terms of $\epsilon$. First, the integrand of the kinetic energy is expanded to obtain
\begin{align} \label{eq:kinetic_energy_expand}
\mathcal{E}_K =  \iint_{\epsilon h}^1 \frac{U^2}{2} dz dx + \epsilon \iint_{\epsilon h}^1 U u_1 dz dx + \epsilon^2 \iint_{\epsilon h}^1 \left( \frac{u_1^2 + w_1^2}{2} + U u_2 \right) dz dx + \mathcal{O}(\epsilon^3) ,
\end{align}
where $u_1 = - \partial_z \psi_1$, $w_1 = \partial_x \psi_1$, and $u_2 = - \partial_z \psi_2$ are introduced.
To remove the dependence on $\epsilon$ of the integration interval, we use a formula valid for an arbitrary analytic function $f(z)$,
\begin{align}
\int_{\epsilon h}^1 f dz = \int_0^1 \left[ f - \delta(z) \left( \epsilon h f + \frac{\epsilon^2 h^2 f_z}{2} + \ldots \right) \right] dz ,
\end{align}
where $\delta$ is the Dirac delta function. Consequently, the kinetic energy \eqref{eq:kinetic_energy_expand} is represented as the integration over a rectified domain as
\begin{align} \label{eq:energy_expand_2}
\mathcal{E}_K & = \int_0^1 \frac{U^2 \left< 1 - \epsilon \delta(z) h \right>}{2} dz + \epsilon\int_0^1 U \left< u_1 \right> dz \nonumber \\
& + \epsilon^2 \int_0^1 \left( \left< \frac{u_1^2 + w_1^2}{2} \right> + U \left< u_2^\ast \right> \right) dz + \mathcal{O}(\epsilon^3) ,
\end{align}
in which the horizontal integration is denoted by $\left< \ \right> \equiv \int dx$, and the second-order horizontal velocity is redefined as
\begin{align} \label{eq:u2_delta}
u_2^\ast \equiv u_2 - \delta (z) \left( h u_1 + \frac{h ^2 U_z}{2} \right) .
\end{align}
Note that the additional terms represent the volume transport associated with the corrugation in the bottom boundary. If $\left< h \right>$ were to vanish, the horizontal average of these terms would correspond to the so-called bolus velocity\textemdash a residual transport induced by the correlation between velocity and fluid thickness\footnote{This kind of transport for topography-generated internal waves is hinted in the footnote on page~135 of \cite{buhler2014waves}.}.

The first term on the right-hand side of \eqref{eq:energy_expand_2} does not depend on time and is unimportant. To evaluate the second term, we shall integrate \eqref{eq:first_order_equation_psi} over $x$ to obtain $\partial_t \partial_z \left< u_1 \right> = 0$, i.e., $\partial_z \left< u_1 \right>$ is constant in time. \yo{It is natural to} choose the initial condition such that $\partial_z \left< u_1 \right> = 0$, which yields, from \eqref{eq:first_order_boundary_condtion}, $\left< u_1 \right> = \left[ U(0) - U_T \cos \omega_T t \right] \left< h \right>$. We thus understand that the second term on the right-hand side of \eqref{eq:energy_expand_2} is also determined by the reference state and irrelevant to the wave dynamics. It is enough to concentrate our attention on the remaining $\mathcal{O}(\epsilon^2)$ terms.

In physical oceanography, it is \yo{widely} known \citep[e.g.,][]{vallis2017atmospheric} that the potential energy in \eqref{eq:totak_potential_energy} can be separated into a constant part and a varying part, referred to as the background potential energy and the available potential energy, respectively.
We present its detailed derivation in Appendix~\ref{sec:app_energy} and write the final result here as
\begin{align} \label{eq:available_potential_energy}
\mathcal{E}_{P} = \mathcal{E}_{BP} + \epsilon^2 \int_0^1 \left< \frac{\rho^2_1}{2 N^2} \right> dz + \mathcal{O}( \epsilon^3 ) ,
\end{align}
where $\mathcal{E}_{BP}$ is a constant.

Combining \eqref{eq:kinetic_energy_expand}, \eqref{eq:energy_expand_2} and \eqref{eq:available_potential_energy}, we represent the total amount of energy in a single expression as
\begin{align} \label{eq:energy_sum}
\mathcal{E}_K + \mathcal{E}_P = \mathcal{E}_B + \epsilon^2 \int_0^1 \left( \left< E \right> + U \left< u_2^\ast \right> \right) dz + \mathcal{O}(\epsilon^3) ,
\end{align}
where $\mathcal{E}_B$ is an unimportant part and can be ignored. In the remaining terms, we have defined
\begin{align}
E \equiv \frac{u_1^2 + w_1^2}{2} + \frac{\rho_1^2}{2N^2} ,
\end{align}
which is a quadratic function of the $\epsilon$-order disturbance amplitude and the common definition of the energy density of internal gravity waves. The other part in \eqref{eq:energy_sum}, $\int_0^1 U \left< u_2^\ast \right> dz$, belongs to the energy contained in the horizontal mean flow. Variation in this energy can be regarded as feedback to the mean flow from the deviations. Importantly, wave energy $E$ is not conserved in shear flow, but, as follows from its definition, the summation of the two terms of disturbance energy, $\left< E \right> + U \left< u_2^\ast \right>$, should be conserved in the absence of an external force.

We shall investigate the energy variations in the system induced by the topographic forcing. First, multiplying $-\psi_1$ to \eqref{eq:first_order_equation_psi} and $\rho_1 / N^2$ to \eqref{eq:first_order_equation_rho}, combining them, and performing the horizontal integration, we derive
\begin{align}
\frac{\partial \left< E \right>}{\partial t} + \frac{\partial}{\partial z} \left< - \psi_1 \frac{\partial^2 \psi_1}{\partial t \partial z} \right> = U \frac{\partial \left< u_1 w_1 \right>}{\partial z} . \label{eq:energy_mean_E}
\end{align}
In this expression, $\left< u_1 w_1 \right>$ appearing on the right-hand side stands for the Reynolds stress, i.e., the vertical flux of the horizontal momentum. Through this term, waves exchange energy with the horizontal mean flow. Next, performing the horizontal integration to the horizontal part of equations in \eqref{eq:equation_u_prime} and collecting the $\epsilon^2$-order terms, we derive an equation governing the variations in the mean flow as
\begin{align} \label{eq:mean_u_2}
\frac{\partial \left< u_2 \right>}{\partial t} + \frac{\partial \left< u_1 w_1 \right>}{\partial z} = P .
\end{align}
Here, $P$ denotes the difference in the pressure $p_2$ between $x = - \infty$ and $x = \infty$ and is uniform in the vertical direction; because the fluid is motionless sufficiently far from the localized \yo{topography}, a hydrostatic balance is maintained there, and the pressure deviation from the reference state does not depend on the depth. This pressure force is essential to fulfill the constraint of the net volume transport. That is, expanding \eqref{eq:volume_transport} in terms of $\epsilon$ and collecting the second-order terms, we derive $\int u_2^\ast dz = 0$, which combined with \eqref{eq:mean_u_2} yields
\begin{align} \label{eq:momentum_balance}
\left( \left< u_1 w_1 \right> - \frac{d}{dt} \left<h u_1\right> \right)_{z=0} + P = 0 ,
\end{align}
where we have used $d \left< h^2 \right> / dt = 0$. This expression represents the horizontal momentum balance in the whole system; the first two terms represent the form stress exerted by the bottom topography, which is compensated by the ambient pressure gradient to keep the net volume transport constant.

Even though the barotropic transport of the mean flow is prescribed by the reference state, the baroclinic part of the mean flow can be accelerated/decelerated by the convergence of the Reynolds stress as \eqref{eq:mean_u_2}, and its reaction performs work on wave motion through the right-hand side of \eqref{eq:energy_mean_E}. Combining the two equations in \eqref{eq:energy_mean_E} and \eqref{eq:mean_u_2}, we may describe the energy balance at each depth as
\begin{align} \label{eq:energy_equation_flux}
\frac{\partial \left( \left< E \right> + U \left< u_2 \right> \right)}{\partial t} + \frac{\partial}{\partial z} \left< - \psi_1 \frac{\partial^2 \psi_1}{\partial t \partial z} \right> & = PU .
\end{align}
The physical interpretation of the second term on the left-hand side is drawn from the horizontal momentum equation, $- \partial_t \partial_z \psi_1 = \partial_t u_1 = - U \partial_x u_1 - U_z w_1 - \partial_x p_1$, and the integration by parts; this term represents the divergence of energy flux composed of pressure work and kinetic energy advection, i.e., 
\begin{align} \label{eq:pressure_Reynolds}
\left< - \psi_1 \frac{\partial^2 \psi_1}{\partial t \partial z} \right> = \left< p_1 w_1 \right> + \left< U u_1 w_1 \right> .
\end{align}
Vertically integrating \eqref{eq:energy_equation_flux}, using \eqref{eq:momentum_balance}, and taking into account the correction term at the bottom in \eqref{eq:u2_delta}, we write down the net production rate of disturbance energy as
\begin{align} \label{eq:energy_production_define}
\frac{d}{dt} \int_0^1 \left( \left< E \right> + U \left< u^\ast_2 \right> \right) dz = \mathcal{P}^{(T)} + \mathcal{P}^{(S)} + \frac{d}{dt} \left( \left( U - \overline{U} \right) \left< h \frac{\partial \psi_1}{\partial z} \right> \right)_{z = 0} \equiv \mathcal{P}(t)
\end{align}
with
\begin{align} \label{eq:energy_production_TS}
\mathcal{P}^{(T)} \equiv \left< - \psi_1 \frac{\partial^2 \psi_1}{\partial t \partial z} \right>_{z = 0} \quad \text{and} \quad
\mathcal{P}^{(S)} \equiv - \overline{U} \left< u_1 w_1 \right>_{z=0} ,
\end{align}
where we have written the barotropic velocity as $\overline{U} \equiv \int_0^1 U dz$. The third term in the middle expression of \eqref{eq:energy_production_define} represents the variations in the kinetic energy associated with the volume transport on the varying boundary. This term owes its existence to the vertical gradient of the steady flow; if $U$ is uniform, $U_{z=0} - \overline{U}$ identically vanishes. In this form, the difference between the barotropic velocity and the velocity at the bottom demands a minor modification in the energy budget of the whole system. In any case, a long-term average of this term inevitably vanishes for a solution whose amplitude is bounded all the time. Therefore, volume transport induced by the corrugated topography makes no contributions to the net energy conversion from barotropic to baroclinic components. In the remaining terms of \eqref{eq:energy_production_define}, as elucidated in \eqref{eq:pressure_Reynolds}, $\mathcal{P}^{(T)}$ stands for the vertical energy flux from the bottom and accounts for the energy supply from the oscillatory barotropic tide. On the other hand, $\mathcal{P}^{(S)}$ corresponds to the pressure work done by the steady flow.

\subsection{\yo{Pseudomomentum and pseudoenergy}} \label{sec:pseudomomentum}
As we have seen, the second-order horizontal velocity, $u^\ast_2$, plays a key role in the energetics. Unlike the other first-order quantities, this variable is difficult to directly obtain. Nonetheless, its horizontal integration, $\left< u^\ast_2 \right>$, can be \yo{diagnosed} based on the other variables by use of the concept of pseudomomentum. Adding the bottom-correction term to that formulated by \cite{shepherd1990symmetries}, let us define the pseudomomentum density\footnote{The present definition of $\textsf{p}$ involves only Eulerian variables, and it differs from the pseudomomentum density in the generalized Lagrangian mean theory. For a plane wave limit, averaged over a phase, the two definitions will coincide \citep{buhler2014waves}.} as
\begin{align} \label{eq:pseudomomentum_density}
\textsf{p} = \frac{\rho_1 \nabla^2 \psi_1}{N^2} - \frac{U_{zz} \rho_1^2}{2 N^4} - \delta (z) \left( h u_1 + \frac{h ^2 U_z}{2} \right) ,
\end{align}
which is a quadratic quantity in terms of $(h, \psi_1, \rho_1)$. The horizontal integration of $\textsf{p}$ follows a similar form of the equation as for $\left< u_2^\ast \right>$; specifically, from \eqref{eq:first_order_equation},
\begin{align} \label{eq:pseudomomentum_flux}
\frac{\partial \left< \textsf{p} \right>}{\partial t} + \frac{\partial \left< u_1 w_1 \right>}{\partial z} = 0    
\end{align}
for $z \neq 0$ is established. Comparing this to \eqref{eq:mean_u_2}, we learn that the two equations differ only by the vertically uniform pressure force $P$. Given that $P$ is determined from a constraint $\int_0^1 u_2^\ast dz = 0$, we come up with a relevant quantity for the pseudomomentum by projecting $\textsf{p}$ to the baroclinic component, i.e., $\textsf{p} - \int_0^1  \textsf{p} dz \equiv \textsf{p}'$. Consequently, $\left< \textsf{p}' \right>$ is shown to obey the same equation as $\left< u_2^\ast \right>$. Therefore, neglecting the difference in their initial conditions, the $\epsilon^2$-order modification in the horizontal mean velocity, $\left< u_2^\ast \right>$, is identified as the baroclinic part of the pseudomomentum, $\left< \textsf{p}' \right>$.

In addition to the pseudomomentum density, it is also informative to define the two kinds of pseudoenergy density as $\textsf{e} = E + U \textsf{p}$ and $\textsf{e}' = E + U \textsf{p}'$. Note that the original definitions $\textsf{p}$ and $\textsf{e}$ are the locally conserved quantities\textemdash obeying equations \eqref{eq:pseudomomentum_flux} and
\begin{align} \label{eq:pseudoenergy_flux}
\frac{\partial \left< \textsf{e} \right>}{\partial t} + \frac{\partial}{\partial z} \left< - \psi_1 \frac{\partial^2 \psi_1}{\partial t \partial z} \right> = 0
\end{align}
for $z \neq 0$. However, to discuss the production rate of the disturbance energy, the modified form of pseudoenergy $\textsf{e}'$ is more relevant because its volume integral coincides with the $\epsilon^2$-order disturbance energy defined in \eqref{eq:energy_sum}. It is enlightening to write the equivalence of three forms of energy representation,
\begin{align} \label{eq:energy_three_forms}
\int_0^1 \left< \textsf{e}' \right> dz = \int_0^1 \left< E \right> dz + \int_0^1 U \left< u_2^\ast \right> dz = \int_0^1 \left< \textsf{e} \right> dz - \overline{U} \int_0^1 \left< \textsf{p} \right> dz .
\end{align}
The final expression as well as \eqref{eq:pseudomomentum_flux} and \eqref{eq:pseudoenergy_flux} provide another physical interpretation for \eqref{eq:energy_production_define} and \eqref{eq:energy_production_TS}; $\mathcal{P}^{(T)}$ and $\mathcal{P}^{(S)}$ stand for the vertical flux of pseudoenergy and pseudomomentum multiplied by minus the barotropic velocity, both of which are injected at the bottom, respectively. The disturbance energy generated on the bottom at the rate of $\mathcal{P}$ is partitioned into the wave part, $\int_0^1 \left< E \right> dz$, and the mean flow part, $\int_0^1 U \left< \textsf{p}' \right> dz$. The mean flow is then accelerated at a rate of $\partial_t \left< u_2^\ast \right> = \partial_t \left< \textsf{p}' \right>$ at each depth via the convergence of the Reynolds stress and the barotropic pressure force. Even though the volume integral of $\textsf{p}'$ is identically $0$, this quantity plays the role of vertical redistribution of momentum through the Reynolds stresses.

Before concluding this subsection, we highlight a marked contrast in energetics between two typical scenarios of topographic wave generation with distinct horizontal boundary conditions: the unbounded system analyzed in this study versus a system with periodic boundary conditions. In the atmospheric context, the periodic conditions are more relevant, as the primary focus is often on the interactions between bottom-generated waves and a planetary-scale zonal mean flow. Under such conditions, the pressure gradient integrated across the zonal direction vanishes, leading to variations in the total volume transport in response to momentum injection at the bottom boundary. The acceleration rate of the mean flow at each altitude can be evaluated using the temporal evolution of the zonally integrated pseudomomentum, $\left< \textsf{p} \right>$. Similarly, variations in the total energy correspond directly to changes in the pseudoenergy, $\int \left< \textsf{e} \right> dz$. However, if the bottom topography is stationary (i.e., in the absence of oscillatory barotropic forcing), the amount of pseudoenergy for bottom-generated waves is null. It is because, in such a case, the energy conversion from the steady current to wave motion via topographic interactions represents an internal redistribution of energy within the coupled wave-mean flow system. Consequently, this conversion is not reflected in the pseudoenergy production rate. This outcome is a well-established result in lee wave generation theory \citep[e.g.,][page 136]{buhler2014waves}. In contrast, in a horizontally unbounded system, as considered in this study, the energetic framework fundamentally differs due to the presence of a net pressure force acting far from the bottom topography. \yo{That is to say}, a steady flow can sustain the generation of internal waves without losing its energy, supported by an external pressure force applied to the system.

\subsection{\yo{Energy conversion rate for a stationary state}} \label{sec:energy_production}
We now examine the disturbance energy production rate, as specified by \eqref{eq:energy_production_define} and \eqref{eq:energy_production_TS}, for a stationary state \yo{considered in Section~\ref{sec:stationary_solution}}. Given that all variables are periodic with period $T \equiv 2 \pi / \omega_T$, the net energy production rate is quantified by averaging over time. The disturbance motions conceptually involve both the barotropic and baroclinic components. However, since the barotropic velocity is prescribed by the reference state, its energy remains bounded. Consequently, the time-averaged energy production rate essentially represents the conversion rate from the barotropic energy in the reference state to the baroclinic energy in disturbances.

To inspect the contribution from the steady flow part $\mathcal{P}^{(S)}$, we first focus on the topographic form stress, which is equivalent to the pseudomomentum input or minus the ambient pressure force, as shown in \eqref{eq:momentum_balance}. Using Parseval's theorem, we may write the time-averaged form stress as
\begin{align}
\frac{1}{T}\int_0^T (-P) dt = \frac{1}{T} \int_0^T \left< u_1 w_1 \right>_{z=0} dt = \sum_{n = -\infty}^\infty \tau_n  .
\end{align}
An explicit expression for the form stress of the $n$th harmonic component is found by inserting the modal expansion \eqref{eq:Fourier_transform} and \eqref{eq:solution_all} in $\left< u_1 w_1 \right>$ with a condition $\hat{\psi}_{1n}(z; k, r) = \hat{\psi}^\dag_{1-n}(z; - k, r)$, leading to
\begin{align} \label{eq:form_stress_n}
\tau_n & = \lim_{r \to 0} \Re \left[ \frac{\ii}{2\pi} \int k \hat{\psi}^\dag_{1n} \frac{\partial \hat{\psi}_{1n}}{\partial z} d k \right]_{z=0}  \nonumber \\
& = \lim_{r \to 0} \frac{1}{2 \pi}  \int k \left\vert \hat{\psi}_{1n}(0; k) \right\vert^2 \Im \left[ \frac{- A_{1z}(0; k, \omega_n + \ii r)}{D(k, \omega_n + \ii r)} \right] dk ,
\end{align}
where the second equality stems from the definition of $\hat{\psi}_{1n}$ in (\ref{eq:solution_single}), its bottom condition is $\hat{\psi}_{1n}(0; k) = (k U_0 - \omega_n) \hat{h}(k) J_n(k U_T / \omega_T) / k$, $A_1$ is defined in \eqref{eq:A01}, and $D(k, \sigma)= A_1(0; k, \sigma)$ is understood. The integrand is finite when $\omega_n$ belongs to the continuous part of the spectrum $\Sigma^c_k$, and we shall invert this condition to write $k \in K^c_{\omega_n}$. For $k \notin K^c_{\omega_n}$, on the other hand, since $A_1$ is real in the limit of $r \to 0$, the integrand vanishes almost everywhere. The exceptional points are the zeros of $D$, i.e., the discrete parts of the spectrum, $K^d_{\omega_n}$. Using \eqref{eq:D_expand}, contributions from these points can be evaluated by using the formula
\begin{align}
\lim_{r \to 0} \Im \frac{- 1}{2 \pi D(k, \sigma + \ii r)} = \sum_j \frac{\delta(k - k_j)}{2 \left\vert c_j^g (\sigma) \right\vert D_\sigma (k, \sigma)} ,
\end{align}
which is valid for $k \notin K^c_\sigma$. Here, we have used
\begin{align*}
D(k, \sigma + \ii r) & = D(k, \sigma) + \ii r D_\sigma(k, \sigma) + \mathcal{O}(r^2)  \\
& = (k - k_j) F(k, \sigma) + \ii r D_\sigma(k, \sigma) + \mathcal{O}(r^2) 
\end{align*}
with $F = D_k$ at $k = k_j$ understood and a formula of a nascent delta function
\begin{align*}
\lim_{r\rightarrow \pm 0} \frac{r}{\pi (x^2+r^2)} = \pm \delta(x) .
\end{align*}
The form stress for the $n$th harmonic component \eqref{eq:form_stress_n} is consequently represented as
\begin{align}
\tau_n & = \sum_{k_j \in K^d_{\omega_n}} k_j \chi_j (\omega_n) \left\vert \hat{\psi}_{1n}(0; k_j) \right\vert^2 + \frac{1}{2 \pi} \int_{K^c_{\omega_n}} k R(k, \omega_n) \left\vert \hat{\psi}_{1n}(0; k) \right\vert^2 dk
\end{align}
with
\begin{align} \label{eq:coefficient}
\chi_j (\sigma) \equiv \frac{A_{1z}(0; k_j, \sigma)}{2 \left\vert c_j^g(\sigma) \right\vert D_\sigma(k_j, \sigma)} \quad \text{and} \quad R(k, \sigma) \equiv - \Im \left[ \frac{ A_{1z}(0; k, \sigma)}{D(k, \sigma)} \right] .
\end{align}
The two functions $\chi_j$ and $R$ play the roles of transforming a prescribed boundary condition of $\psi$ at $z=0$ to the vertical wave action flux at the bottom. Here, wave action is regarded as the pseudomomentum divided by the wavenumber. A more common definition of the wave action would be the wave energy divided by the intrinsic frequency\textemdash wave frequency observed in a frame moving with a background current. Indeed, this classical definition also applies to the present case, but care should be taken that the background current is not uniform. This point will be discussed in Section~\ref{sec:discussion}. We shall call $\chi_j$ and $R$ the action production coefficients for the discrete and continuous spectra, respectively.

\yo{
With the harmonic decomposition of the form stress in hand, the time-averaged energy conversion rate from the steady background flow to the baroclinic motion follows directly as
$(1/T) \int_0^T \mathcal{P}^{(S)} \, dt = -\overline{U} \sum_{n = -\infty}^\infty \tau_n$.
Similarly, the time-averaged energy conversion rate from the barotropic tide to the baroclinic motion, or equivalently the pseudoenergy production rate, is written as
$(1/T) \int_0^T \mathcal{P}^{(T)} \, dt = \sum_{n = -\infty}^\infty \mathcal{P}^{(T)}_n$ with
\begin{align} \label{eq:energy_production_T}
\mathcal{P}^{(T)}_n = \sum_{k_j \in K^d_{\omega_n}} \omega_n \chi_j (\omega_n) \left\vert \hat{\psi}_{1n}(0; k_j) \right\vert^2 + \frac{1}{2 \pi} \int_{K^c_{\omega_n}} \omega_n R(k, \omega_n) \left\vert \hat{\psi}_{1n}(0; k) \right\vert^2 dk .
\end{align}
This expression makes clear that the pseudoenergy flux is given by the wave frequency multiplied by the wave-action flux, consistent with the standard pseudoenergy--action relation. In particular, the pseudoenergy production rate of the zero-frequency component vanishes identically, $\mathcal{P}^{(T)}_0 = 0$; for this mode, the baroclinic motion is driven solely by the steady background flow, and the oscillatory tidal forcing does not contribute energetically, a general consequence of the pseudoenergy-based framework \citep{buhler2014waves}. On the other hand, the form stress can remain finite even for the zero-frequency component.

Combining the steady and tidal contributions, we obtain the total time-averaged barotropic-to-baroclinic energy conversion rate,
$(1/T) \int_0^T \mathcal{P} \, dt = \sum_{n = -\infty}^\infty \mathcal{P}_n$, with
\begin{align}
\mathcal{P}_n & = \sum_{k_j \in K^d_{\omega_n}} \left( \omega_n - \overline{U} k_j \right) \chi_j (\omega_n) \left\vert \hat{\psi}_{1n}(0; k_j) \right\vert^2 \nonumber \\
& + \frac{1}{2 \pi} \int_{K^c_{\omega_n}} \left( \omega_n - \overline{U} k \right) R(k, \omega_n) \left\vert \hat{\psi}_{1n}(0; k) \right\vert^2 dk .  \label{eq:energy_conversion_rate_n}
\end{align}
By analogy with the pseudomomentum or pseudoenergy flux, one might be tempted to interpret \eqref{eq:energy_conversion_rate_n} as a wave-energy flux written as an intrinsic frequency times an action flux. However, this interpretation is generally incorrect because $\omega_n - \overline{U} k$ is not an intrinsic frequency unless $U(z)$ is uniform. It is important to remember that $(1/T)\int_0^T \mathcal{P} \, dt$ represents the barotropic-to-baroclinic energy conversion rate, not the wave-energy production rate.

Returning to \eqref{eq:coefficient}, the discrete-spectrum coefficient $\chi_j$ depends on the group velocity $c_j^g$. If $c_j^g$ happens to vanish, $\chi_j$ diverges and the form stress and the conversion rate become unbounded within inviscid linear theory. Aside from this special case, the time-averaged conversion rate is finite and time independent. In an unbounded domain, the injected disturbance energy is carried away by radiating motions toward $x=\pm\infty$; consequently, the total disturbance energy integrated over the domain grows linearly in time.

This constant-rate growth can be interpreted as a quasi-resonant response of a spatially unbounded system to spatially localized forcing. As discussed in Section~\ref{sec:spectrum}, for each fixed horizontal wavenumber $k$ the Taylor--Goldstein equation admits a discrete set of regular eigenfrequencies, say $\sigma=\sigma_j(k)$. As $k$ varies, these eigenfrequencies depend continuously on $k$, forming dispersion curves in the $(k,\sigma)$ plane. Because localized topography excites a continuum of horizontal wavenumbers, a given forcing frequency $\omega_n$ can interact with these dispersion curves at isolated wavenumbers $k_j$ satisfying $D(k_j,\omega_n)=0$ (i.e.\ $k_j\in K^d_{\omega_n}$). When the intersection is regular (non-vanishing group velocity), each contribution remains finite and the forcing supplies energy at a constant rate. \cite{de2020attractors} referred to this mechanism as quasi-resonance and contrasted it with a genuine resonance, in which the energy grows faster than linearly so that a time-independent conversion rate is not defined. The divergence of $\chi_j$ at $c_j^g=0$ corresponds to the onset of this resonant behaviour.
}

\yo{
To illustrate the conversion-rate formula \eqref{eq:energy_conversion_rate_n}, Fig.~\ref{fig:energy_conversion_rates} shows numerical evaluations for a simple background state. We take uniform stratification $N=1$ and a linear shear flow $U(z)=0.05+0.15z$, together with a barotropic tide of amplitude $U_T=0.2$ and frequency $\omega_T=0.2$. The topography is chosen as an extremely narrow ridge; accordingly, its Fourier transform is nearly flat, and for simplicity we set $\hat{h}(k)=1$ in the computation. We evaluate separately the contributions associated with the tidal and steady-current parts and further split each into the discrete- and continuous-spectrum contributions. Harmonics up to $n=2$ are retained ($n=0,1,2$). For the discrete spectrum we also separate left- and right-propagating contributions according to the sign of the horizontal group velocity $c^g$. In this example, a substantial contribution comes from the discrete modes of the zero-frequency component ($n=0$), i.e.\ lee waves. For the tide-frequency component ($n=1$), the upstream-propagating branch ($c^g<0$) is stronger than the downstream-propagating branch ($c^g>0$), consistent with the numerical simulations of \cite{lamb2018internal}.

\begin{figure}[t]
\centering\includegraphics[width=\columnwidth]{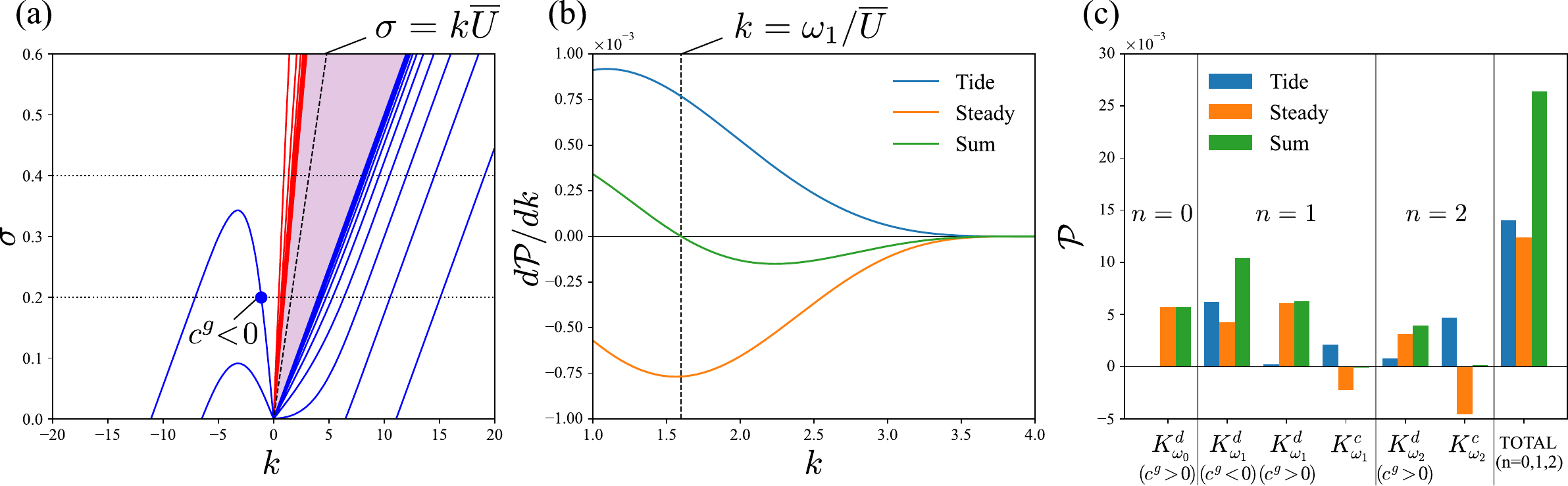}
\caption{\label{fig:energy_conversion_rates} 
\yo{Numerically computed energy conversion rates for the parameter set described in the text ($N=1$, $U(z)=0.05+0.15z$, $U_T=0.2$, $\omega_T=0.2$, and $\hat{h}(k)=1$). (a) Spectrum of the operator $\mathsfbi{M}_k$: red and blue curves show the discrete spectrum (dispersion curves of regular eigenmodes), while the magenta shading indicates the continuous spectrum. The dashed line corresponds to $\sigma/k=\overline{U}=0.125$ (phase speed equal to the barotropic velocity). Horizontal dotted lines indicate the forcing frequencies $\sigma = \omega_n$ for $n=1,2$; the blue dot marks an upstream-propagating intersection with the discrete spectrum ($c^g<0$). (b) Integrand of the continuous-spectrum contribution in \eqref{eq:energy_conversion_rate_n} for $n=1$, separated into the tidal part, the steady-current part, and their sum. The vertical dashed line indicates $k=\omega_1/\overline{U}$. (c) Contributions to the conversion rates from the discrete and continuous spectra for $n=0,1,2$. Discrete-spectrum contributions are further divided into left- ($c^g<0$) and right-propagating ($c^g>0$) components. For $n=1,2 $, the plotted values are multiplied by 2 to include the equal contributions from the negative harmonics $n=-1, -2$ (i.e. $\mathcal{P}_{-n}=\mathcal{P}_n$).}}
\end{figure}

Although the net contribution from the continuous spectrum is small in this example, this does not imply that the critical-level (singular) response is negligible. Rather, the tidal and steady-current parts contribute comparably but with opposite signs, leading to a strong cancellation. This is illustrated in Fig.~\ref{fig:energy_conversion_rates}b, which plots the integrand of the continuous-spectrum contribution to \eqref{eq:energy_conversion_rate_n} for $n=1$. Over the integration interval, the tidal part (proportional to $\omega_n$) is positive whereas the steady-current part (proportional to $-\overline{U}k$) is negative, because $R(k,\omega_n)>0$. Their magnitudes cross at $k=\omega_n/\overline{U}$ (vertical dashed line), where the sum changes sign. For the second harmonic ($n=2$), the discrete-spectrum contribution is smaller and the continuous-spectrum contribution becomes more prominent; although the two parts still oppose each other, their individual magnitudes are larger than in the $n=1$ case. This trend suggests that continuous-spectrum effects become increasingly important for higher-frequency forcing.
Notably, in classical no-shear formulations \citep[e.g.,][]{khatiwala2003generation}, the spectrum is purely discrete and no continuous-spectrum (critical-layer) contribution arises.
The present results suggest that this discrete-mode-only approximation can miss an important part of the conversion budget when background shear is present, especially at higher harmonics.
}

\yo{
In the computations above, $A_1$ and $D$ in \eqref{eq:A01} and \eqref{eq:dispersion} were evaluated using the closed-form expressions derived in Appendix~\ref{sec:app_uniform_shear}. For the continuous spectrum, the calculation can be streamlined by expressing $A_1(z;k,\sigma)$ in terms of a pair of linearly independent solutions $\gamma^\pm(z;k,\sigma)$ of the Taylor--Goldstein equation (Appendix~\ref{sec:app_spec}). When the Richardson number at the critical level is large, $N^2/U_z^2\gg 1/4$ at $z=z^c$, the connection rule \eqref{eq:connection_rule} implies $|\gamma^+(0)|\gg|\gamma^-(0)|$, whereas $|\gamma^+(1)|=|\gamma^-(1)|$.
Accordingly, the action production coefficient can be approximated by
\begin{align}
R(k,\sigma) \approx -\Im\left[\frac{\gamma^+_z(0)}{\gamma^+(0)}\right] \equiv R_\infty(k,\sigma),
\end{align}
which is essentially insensitive to the upper boundary condition. Physically, when the Richardson number at the critical level is large, incident waves are mostly absorbed and do not reflect back toward the bottom, so the problem approaches the classical unbounded setting. Indeed, setting $U_0=0$ and applying a WKB approximation near $z=0$ yields $\gamma^+ \propto e^{\ii m z}$ with $m=-|k|\sqrt{N^2-\sigma^2}/\sigma$. One then finds $\sigma R_\infty \approx |k|\sqrt{N^2-\sigma^2}$ for $|\sigma|<N$, so that the continuous-spectrum term in \eqref{eq:energy_production_T} reduces to equation (5.3) of \cite{bell1975lee}. A similar insensitivity to the upper boundary has been noted in finite-depth viscous problems when strong dissipation prevents reflection \citep{shakespeare2017viscous,shakespeare2021dissipating}. In oceanographic applications, Bell's unbounded formula is often used to estimate the generation of short-wavelength lee waves \citep{legg2021mixing}; the present analysis provides a rationale for this practice. Nevertheless, whether turbulent mixing can be parameterized based solely on such elementary lee-wave theory remains debated, as discussed in the next section.
}

\section{Discussion} \label{sec:discussion}

\begin{figure}[t]
\centering\includegraphics[width=0.68\columnwidth]{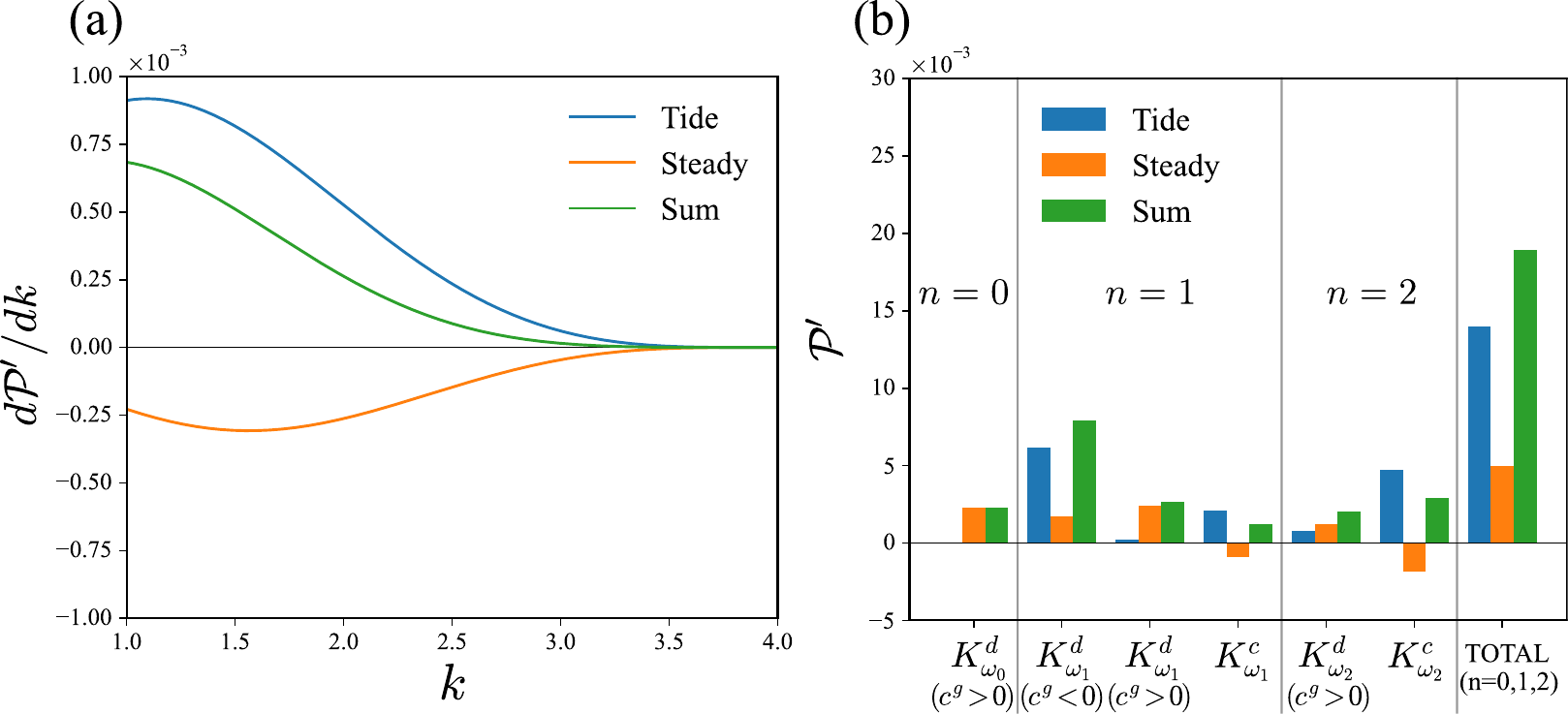}
\caption{\label{fig:modified_energy_conversion_rates} \yo{Numerically computed modified conversion rates $\mathcal{P}'$ defined by \eqref{eq:energy_conversion_rate_modified} averaged over time, for the same parameter setting as in Fig.~\ref{fig:energy_conversion_rates}. (a) Integrand of the continuous-spectrum contribution for $n=1$, separated into the tidal part, the steady-flow part, and their sum. (b) Contributions to the conversion rates from the discrete and continuous spectra for $n=0,1,2$.}}
\end{figure}

The central result in the present study is the explicit formula for the barotropic-to-baroclinic energy conversion rate \eqref{eq:energy_conversion_rate_n}, which warrants further discussion in light of similar formulas found in the literature. In previous studies that addressed internal wave generation over a seafloor topography forced by vertically uniform flow \citep[e.g.,][]{khatiwala2003generation,nikurashin2010radiation}, the energy conversion rate was shown to equal the energy flux composed as a product of pressure and vertical velocity at the bottom. If the reference flow $U(z)$ is uniform, our formulation is in line with this result; in such a special situation, as inferred from \eqref{eq:pressure_Reynolds} and \eqref{eq:energy_production_TS}, the Reynolds stress terms in $\mathcal{P}$ are canceled out so that $\mathcal{P} = \left< p_1 w_1 \right>_{z=0}$ is recovered. For general cases with $\overline{U} \neq U_0$, on the other hand, the barotropic-to-baroclinic energy conversion rate $\mathcal{P}$ does not coincide with the vertical pressure work. This discrepancy would be reconciled by replacing the barotropic velocity $\overline{U}$ in \eqref{eq:energy_production_define} with the velocity at the bottom $U_0$ to define a modified form of the energy production rate
\begin{align} \label{eq:energy_conversion_rate_modified}
\mathcal{P}' \equiv \left< - \psi_1 \frac{\partial^2 \psi_1}{\partial t \partial z} \right>_{z = 0} - U_0 \left< u_1 w_1 \right>_{z=0} = \left< p_1 w_1 \right>_{z=0} .
\end{align}
The corresponding time-averaged energy production rate for the $n$th tidal harmonic is obtained by replacing $\overline{U}$ with $U_0$ in \eqref{eq:energy_conversion_rate_n}. \yo{With this choice,} the modified energy production rate coincides with the action flux multiplied by the intrinsic frequency at the bottom, \yo{suggesting an interpretation of $\mathcal{P}'$ as the flux of} wave energy $\left< E \right>$. \yo{Figure~\ref{fig:modified_energy_conversion_rates} demonstrates a modified energy conversion rate $\mathcal{P}'$ (averaged over time) for the same parameter setting as Fig.~\ref{fig:energy_conversion_rates}. Due to the replacement of $\overline{U}$ with $U_0$, the steady flow contribution is reduced. As a result, the continuous spectrum part makes a marked positive contribution, in contrast to the original energy conversion rate $\mathcal{P}$ shown in Fig.~\ref{fig:energy_conversion_rates}c.}

\yo{Energy conversion formulas similar to \eqref{eq:energy_conversion_rate_modified}} have been used by \cite{nikurashin2011global} and \cite{shakespeare2020interdependence} to \yo{estimate} the internal-wave energy radiating away from the seafloor topography forced by geostrophic and tidal flows. However, there remains a concern that the energy production rate defined in \eqref{eq:energy_conversion_rate_modified} depends only on \yo{flow quantities evaluated at the bottom boundary, whereas the present analysis highlights the potential importance of the full vertical structure of the steady shear flow in the energetics of bottom-generated internal gravity waves.} To find the proper definition of the energy conversion rate in shear flow, we should keep in mind that the central motivation for this task is to quantify the amount of energy available for vertical mixing. From this viewpoint, defining the energy production rate in reference to the velocity at the bottom can lead to some errors as the waves exchange the energy with the steady flow during the vertical propagation. \cite{kunze2019energy} clarified this point and argued that the energy exchange between waves and mean flows might explain the significant overestimation of the energy dissipation rates in the Southern Ocean predicted by the conventional formula of the lee wave theory. For the accurate estimation of the wave energy available for mixing, the best definition of the energy production rate would be the one that replaces $\overline{U}$ in \eqref{eq:energy_production_define} with the velocity at a level where wave dissipation takes place, though identifying this level remains a significant challenge.

In the following, we list other limitations of the current study and discuss possible future directions to address them. First, this study employs a two-dimensional model assuming homogeneity in one horizontal direction. We may readily extend the formulation for a three-dimensional model as far as the reference flow varies only in the vertical direction. In that case, for each horizontal wave vector of the generated waves, projecting the horizontal velocity in the same direction as the wave vector, we may derive a solution in the same manner as the two-dimensional case. Second, we have restricted our study to the case where $U_z$ is positive. This assumption was necessary to perform the Frobenius series expansion of the singular solution around a critical level, with the leading-order term specified by \eqref{eq:gamma_critical}. If there exists a level where $U_z$ changes signs, we need special treatment there. \cite{bouchet2010large} considered this kind of problem for a homogeneous fluid and found an interesting \textit{vorticity depletion} phenomenon. Applying their asymptotic approach to a stratified fluid case remains to be done.

Third, we have assumed that the topographic height is small to derive a set of linear equations. In fact, even for a finite-size topography, as far as the tidal excursion is small (i.e., the higher tidal harmonics are negligible), we may derive a linear equation system. A number of previous works solved that problem in the absence of a background flow using the Green function method \citep[e.g.,][]{petrelis2006tidal,echeverri2010internal} or, more recently, a coupled-mode system (CMS) \citep{papoutsellis2023internal}. 
In both approaches, previous studies expanded a solution into a countable number of regular functions, each of which solves a Sturm–Liouville equation derived from an elementary problem with flat-bottom topography. However, once a sheared reference flow is included, a solution can no longer be represented as a superposition of regular functions.
In the special case of constant shear and stratification, \cite{engevik1971note} proved that any solution can be represented as the combination of the sum over a countable set of discrete functions and integration over singular functions. Nonetheless, it is uncertain whether one may evaluate the improper integral involved in Engevik's formula efficiently in a computer via an algorithm analogous to the existing Green function or CMS method. Moreover, for a finite-height topography case, a steady shear flow can exist above the top of the topography, and formulas for a constant shear flow problem are not applicable. Wave generation with finite-height topography in the presence of a surface-intensified steady flow is the situation numerically investigated by \cite{masunaga2019strong} to clarify interactions of Kuroshio and tidal flow over a ridge off the south coast of Japan. We point out that tilting wave phases visible at the downstream side in figure~6 of \cite{masunaga2019strong} and also figure~1 of \cite{lamb2018internal} resemble that of Fig.~6 in the present paper. A recent observation study has reported significant energy dissipation downstream of a tall seamount located in Kuroshio and partly ascribed it to the wave-flow interactions \citep{takahashi2024energetic}. Although the actual situations are much more intricate, the critical-layer process for bottom-generated internal waves, as elucidated in this paper, may ubiquitously occur, leading to intense energy dissipation and mixing in the ocean.

Fourth, and most importantly, a challenging problem arises if we consider an actual ocean condition that involves Earth's rotation. Generally, in a rotating stratified fluid, a steady \yo{horizontal flow with vertical shear} is associated with a horizontal density gradient according to the thermal wind balance. The present-style vertical two-dimensional model makes sense only when the vertical gradient of the steady flow, $U_z$, is constant. Eigenmodes in such a sheared rotating stratified fluid were investigated in numerous studies, and it is known that there exist eigenfrequencies possessing positive imaginary parts, i.e., unstable modes, \yo{across a wide range of Richardson numbers} \citep{eady1949long,stone1966non,nakamura1988scale,molemaker2005baroclinic}. \yo{Furthermore, a transient non-modal growth in wave energy can occur more efficiently than modal counterparts \citep{heifetz2003generalized,zemskova2020transient}. In such situations,} tidally forced waves are expected to \yo{be amplified during propagation}, extracting energy from the steady flow and eventually breaking, leading to significant energy dissipation\footnote{\yo{\cite{maitland2025oceanic} developed a Green-function formulation under the assumption of a neutrally stable modal spectrum. Extending such an approach to settings that admit unstable modes, and to time-periodic tidal forcing, remains to be established.}}. If removing the thermal wind balance while retaining the Coriolis effect, we may consider another relevant situation in which a background flow oscillates with the inertial period. Indeed, a recent numerical study of \cite{hibiya2024revisiting} reported prompt dissipation of tidally forced internal waves through the interaction with near-inertial current shear. We consider the generation of internal waves in unstable or unsteady shear flows to be an intriguing topic deserving of analytical investigation in future studies.

\section{Conclusions} \label{sec:conclusions}

To summarize, this study discussed internal wave generation on small bottom topography forced by a barotropic tide in the presence of a vertically sheared steady flow. \yo{The main objective of the current study was to derive an explicit formula for the barotropic-to-baroclinic energy conversion rate. Conventional studies discussing waves in a static reference state represent this energy conversion rate as a summation over a countable number of discrete eigenmodes that compose a complete set of solutions. Once we include a steady shear flow, the discrete eigenmodes are no longer complete, and we need to take into account singular solutions associated with a critical level. To elucidate this point, we applied the Fourier transform in the horizontal direction and focused on a linear operator that characterizes the temporal evolution of an unforced system. The general solution obtained via the Laplace transform involves the resolvent of this operator, whose singularity condition determines the spectrum of the operator. While the ordinary eigenmodes correspond to an infinite number of simple poles, a critical level produces a branch point parameterized by a frequency, as elucidated through the detailed inspection of the Taylor--Goldstein equation (Appendix~\ref{sec:app_spec}). Consequently, the spectrum is decomposed into discrete and continuous parts, both of which contribute to the net energy conversion rate.

We also demonstrated the asymptotic behaviors of the solutions far from a localized bottom topography. The contributions from the discrete spectrum conform to our expectation that they produce standing wave trains with constant amplitude.
On the other hand, contributions from the continuous spectrum produce spatially evolving wave packets, whose amplitude decays while the velocity shear grows algebraically in the horizontal direction. A surprising point is that there is no wave focusing at a particular level, which contrasts sharply with the basic sinusoidal topography case, where wave rays are attracted to a critical level determined by the forcing frequency and the horizontal wavenumber. In the present case, the localized \yo{topography} generates a continuum of Fourier modes, each associated with its own critical level, preventing wave accumulation at any specific altitude. One could reinterpret this phenomenon as wave accumulation occurring at all levels, far from the \yo{topography}.
}

Although we employed a linearized model that governs small amplitude waves to the leading order, we clarified that the next-order component in the velocity plays an essential role in the total energy budget of bottom-generated disturbances. This component represents the feedback from waves to mean flows and can be virtually replaced by the baroclinic part of the pseudomomentum. The pseudomomentum is quadratic with respect to the leading-order wave amplitude and conserved even in the presence of a shear flow. We also introduced the pseudoenergy through a combination of the wave energy and the pseudomomentum multiplied by the reference flow velocity. We then clarified that the net energy conversion rate from barotropic to baroclinic motions coincides with the production rate of pseudoenergy minus the pseudomomentum multiplied by the barotropic velocity.

In the framework of the present inviscid and linear model, the net energy conversion rate is stationary so that the baroclinic energy grows linearly with time, a situation referred to as quasi-resonance. However, there can be an exceptional case when a genuine resonance occurs; if a group velocity of a standing wave mode vanishes by accident, its energy grows superlinearly. There have been a huge number of studies addressing this problem in various contexts by incorporating weak nonlinearity, such as the representative work done by \cite{grimshaw1986resonant}. Extending the current formulation to a finite-amplitude model is left to a future study.

In conclusion, this paper made clear the roles of the continuous spectrum inherent in a steady shear flow in the generation of internal gravity waves by oscillatory forcing. While primarily motivated by oceanic applications, this study lays a solid foundation for exploring wave-mean flow interactions in oceans, atmospheres, and astrophysical flows driven by periodic forcing. A major challenge moving forward is the integration of rotational effects, which significantly influence wave dynamics in these systems.

\begin{bmhead}[Acknowledgements.]
The authors express their gratitude to Anubhab Roy and Patrice Le Gal for their informative comments at the early stage of this work. The authors thank \yo{three anonymous reviewers,} Takashi Ijichi and Ryosuke Nakashima for carefully reading the manuscript and providing helpful comments. \yo{The authors used ChatGPT 5.2 Pro (OpenAI, accessed from a web browser on February 24th, 2026) to assist with improving the clarity and readability of the manuscript. All scientific content and interpretations remain the responsibility of the authors.}
\end{bmhead}

\begin{bmhead}[Funding.]
This study was supported by JSPS Overseas Research Fellowship and KAKENHI Grant Number JP20K14556. This work was supported in part by the Collaborative Research Program of Research Institute for Applied Mechanics, Kyushu University.
\end{bmhead}

\begin{bmhead}[Declaration of interests.]
The authors report no conflict of interest.
\end{bmhead}

\begin{bmhead}[Data availability statement.]
The Python scripts used to generate the results presented in this study are publicly available in a GitHub repository at https://github.com/yonuki-models/tide-internal-wave-shear.git.
\end{bmhead}

%\appendix
\begin{appen}

\section{Resolvent and spectrum of $\mathsfbi{M}_k$} \label{sec:app_spec}
This section inspects the resolvent of the operator $\mathsfbi{M}_k$ in detail to learn the locations of the spectrum and the behavior of solutions around a critical level. Here, we extend the domain of functions with respect to $z$ to the complex plane $\mathbb{C}$ by assuming $N(z)$ and $U(z)$ as analytic functions.
% [This treatment is rationalized based on the Weierstrass approximation theorem.]
This hypothesis is essential for identifying singular points $z$ in the complex plane that arise for a given complex frequency $\sigma$.

\subsection{Resolvent}
To derive the explicit form of the resolvent of $\mathsfbi{M}_k$ defined in \eqref{eq:operator_Mk}, the first step is to consider
\begin{align} \label{eq:sigma_Mk}
\sigma \bm{v} - \mathsfbi{M}_k \bm{v} = \bm{f} ,
\end{align}
where $\sigma$ is a complex scalar, $\bm{f}(z) = (f_1, f_2)$ is a prescribed function, and $\bm{v}(z) = (v_1, v_2)$ is a variable to be determined. An element of the matrix $\mathsfbi{M}_k$ involves an integral operator $\nabla_k^{-2}$ \yo{whose explicit form is given by
\begin{align} \label{eq:inverse_laplacian}
\nabla^{-2}_k A = \int_0^z \frac{\sinh k \zeta \sinh \left[ k (z - 1) \right] A(\zeta)}{k \sinh k} d\zeta + \int_z^1 \frac{\sinh k z \sinh \left[ k (\zeta - 1) \right] A(\zeta)}{k \sinh k} d\zeta .
\end{align}
Introducing a streamfunction-like variable $\phi(z) \equiv \nabla^{-2}_k v_1$, we may transform the vectorial integral equation \eqref{eq:sigma_Mk} into a \yo{scalar} differential equation.} A straightforward manipulation leads to
\begin{align} \label{eq:phi}
\mathcal{T}_{k,\sigma} \phi = - \frac{f_1}{k U - \sigma} + \frac{k f_2}{(k U - \sigma)^2} \equiv F(z) ,
\end{align}
where $\mathcal{T}_{k,\sigma}$ is defined in \eqref{eq:forced_TG_equation}, with the homogeneous boundary conditions $\phi(0) = \phi(1) = 0$. To write down the solution of this equation, we first define two functions $\gamma^\pm(z; k, \sigma)$ that are a pair of linearly independent solutions of a homogeneous Taylor--Goldstein equation, $\mathcal{T}_{k,\sigma} \gamma = 0$, without boundary conditions. Then, two functions $A_0(z; k, \sigma)$ and $A_1(z; k, \sigma)$ are defined as
\begin{subequations} \label{eq:A01}
\begin{align}
A_0 (z) & = \gamma^-(0) \gamma^+(z) - \gamma^+(0) \gamma^-(z) \\
A_1 (z) & = \gamma^-(1) \gamma^+(z) - \gamma^+(1) \gamma^-(z) ,
\end{align}
\end{subequations}
which satisfy $\mathcal{T}_{k,\sigma} A_{0, 1} = 0$ with one-side boundary conditions, $A_0(0) = 0$ and $A_1(1) = 0$. The general solution of \eqref{eq:phi} is then written as
\begin{align} \label{eq:phi_green}
\phi = \frac{A_1(z)}{W} \int_0^z A_0 (\zeta) F(\zeta) d\zeta + \frac{A_0(z)}{W} \int_z^1 A_1 (\zeta) F(\zeta) d\zeta ,
\end{align}
where $W \equiv A_0 A_{1z} - A_{0z} A_1$ is the Wronskian of $A_0$ and $A_1$. Accordingly, $\bm{v}$ is represented in terms of $\bm{f}$ as $\bm{v} = (\sigma \mathsfbi{I} - \mathsfbi{M}_k)^{-1} \bm{f}$ with
\begin{subequations} \label{eq:solution_v12}
\begin{align}
v_1 & = \nabla^2_k \phi = \frac{- k^2 N^2 + k U_{zz} (k U - \sigma)}{(k U - \sigma)^2} \phi - \frac{f_1}{k U - \sigma} + \frac{k f_2}{(k U - \sigma)^2} \\
v_2 & = \frac{k N^2}{k U - \sigma} \phi - \frac{f_2}{k U - \sigma} .
\end{align}
\end{subequations}
A series of expressions, \eqref{eq:phi} - \eqref{eq:solution_v12}, defines the resolvent of $\mathsfbi{M}_k$.

\subsection{Spectrum} \label{sec:app_spec_classification}
The spectrum of $\mathsfbi{M}_k$ is defined by the condition that the norm of $\bm{v}_k$ is unbounded for some $\bm{f}$ that has a finite norm. It seems natural here to choose a norm based on wave energy density as $\lVert \bm{v} \rVert \equiv \int_0^1 (- v_1^\dag \nabla_k^{-2} v_1 + \vert v_2 \vert^2 / N^2) dz = \int_0^1 (k^2 \vert \phi \vert^2 + \vert \phi_z \vert^2 + \vert v_2 \vert^2 / N^2) dz$. We explore the boundedness of $\lVert \bm{v} \rVert$ depending on $\sigma$ for a prescribed $k$. For simplicity, here we assume $k > 0$, but the following consideration immediately applies to $k < 0$ by switching the signs of both $k$ and $\sigma$.

\subsubsection{Discrete part} \label{sec:app_dispersion}
By inspection of \eqref{eq:phi_green}, we first notice that the resolvent is unbounded when the Wronskian $W$ vanishes. We rewrite $W$ in terms of the fundamental solutions $\gamma^\pm$ to decompose it into two factors,
$W = W^\gamma D$, with
\begin{align} \label{eq:dispersion}
W^\gamma \equiv \frac{d \gamma^-}{d z} \gamma^+ - \frac{d \gamma^+}{d z} \gamma^-
\quad \mbox{and} \quad
D\equiv \gamma^- (1) \gamma^+ (0) - \gamma^+ (1) \gamma^- (0) .
\end{align}
Because $\gamma^+$ and $\gamma^-$ are linearly independent, $W^\gamma$ does not vanish. The other factor $D$, which is a function of $k$ and $\sigma$, determines the zeros of $W$. This $D$ is related to $A_{0,1}$ by $D = A_1(0) = - A_0(1)$. If $D(k, \sigma) = 0$, $A_0$ and $A_1$ satisfy both the upper and bottom boundary conditions. Therefore, they become the eigenfunctions of the Taylor--Goldstein equation, and for a prescribed $k$, $\sigma$ is the corresponding eigenfrequency. From the Miles-Howard theorem, all the eigenvalues are real in the present problem. As a consequence, the zeros of $D(k, \sigma)$ for an arbitrary $k$ are located on the real axis.

The eigenvalues of the Taylor--Goldstein equation have upper and lower bounds. This is verified by considering the case when $\vert \sigma \vert \gg 1$. In this limit, we derive $\mathcal{T}_{k,\sigma} \gamma \sim \gamma_{zz} - k^2 \gamma = 0$, and a solution is a combination of exponential functions. Obviously, it cannot satisfy the top and bottom boundary conditions simultaneously.

\subsubsection{Eigenvalues are absent between $kU(0)$ and $kU(1)$}
We discuss the locations of eigenvalues more closely. Let us assume that $\sigma$ is a complex number within and close to a line segment $k U(0) \leq \sigma \leq k U(1)$. In this case, a singular point $z = z^c \in \mathbb{C}$ that satisfies $k U(z^c) = \sigma$ is located in the vicinity of a line segment, $0 \leq z \leq 1$. We shall expand solutions of $\mathcal{T}_{k,\sigma} \gamma^\pm = 0$ around $z^c$ using the Frobenius method, which enables us to choose the two solutions as
\begin{align} \label{eq:gamma_critical}
\gamma^\pm(z) = (z - z^c)^{1/2 \pm \ii \nu} K^\pm(z),
\end{align}
where $\nu = (\left. N^2 / U_z^2 \right\vert_{z = z^c} - 1/4)^{1/2} > 0$, and $K^\pm$(z) are analytic functions satisfying
\begin{align} \label{eq:Gamma}
\frac{d^2 K^\pm}{d z^2} + \frac{1 \pm 2 \ii \nu}{z - z^c} \frac{d K^\pm}{d z} + \left[ \frac{1}{(z - z^c)^2} \left( \frac{N^2}{S^2} - \nu^2 - \frac{1}{4} \right) - \frac{U_{zz}}{(z - z^c) S} - k^2 \right] K^\pm = 0
\end{align}
with $S(z)$ defined by $S = (U - \sigma / k) / (z - z^c)$.

If $z^c$ is real, \eqref{eq:Gamma} is symmetric with respect to taking the complex conjugate and reversing the double signs. Consequently, we can write $K^\pm$ using a single function $K(z)$ as $K^- (z) = K (z)$, $K^+ (z) = \ii \left[ K(z^\dag) \right]^\dag$, where an imaginary factor is placed for convenience so that $A_1$ and $D$ become real functions when $z^c$ is outside $[0, 1]$ as verified from \eqref{eq:A01} and \eqref{eq:dispersion} and the following rule \eqref{eq:connection_rule}. Note that $K(z)$ will not be 0 for any real $z$. Otherwise, $\gamma^\pm = 0$ holds there, contradicting the completeness of the solutions. Because $z = z^c$ is a branch point of \eqref{eq:gamma_critical}, there are two ways to connect the regions $z < z^c$ and $z > z^c$ along the real axis. As we learn from Fig.~\ref{fig:integration_sigma} and \eqref{eq:stationary}, the physically relevant situation is when $\sigma$ approaches the real axis with the imaginary part positive. From the condition $U_z>0$, together with the definition $k U(z^c) =\sigma$, this corresponds to the situation $z^c$ approaches the real axis with the imaginary part positive. Accordingly, we shall choose the Riemann sheet such that
\begin{align} \label{eq:connection_rule}
(z - z^c)^{1/2 \pm \ii \nu} = \begin{cases}
\lvert z - z^c \rvert^{1/2} e^{\pm \ii \nu \log \lvert z - z^c \rvert} & (z - z^c > 0) \\
- \ii e^{\pm \pi \nu} \lvert z - z^c \rvert^{1/2} e^{\pm \ii \nu \log \lvert z - z^c \rvert} & (z - z^c < 0)
\end{cases}
\end{align}
on the real axis. It is convenient to define two functions,
$\tilde{\gamma}^- (z) = \lvert z - z^c \rvert^{1/2 - \ii \nu} K(z), \tilde{\gamma}^+ (z) = \ii \lvert z - z^c \rvert^{1/2 + \ii \nu} \left[ K(z^\dag) \right]^\dag$.

When $0 < z^c < 1$, $D$ cannot be $0$ because
\begin{align}
\lvert D \rvert & = \lvert e^{-\pi \nu} \tilde{\gamma}^- (0) \tilde{\gamma}^+ (1) - e^{\pi \nu} \tilde{\gamma}^+ (0) \tilde{\gamma}^- (1) \rvert \nonumber \\
& > (e^{\pi \nu} - e^{-\pi \nu}) \lvert z^c \rvert \lvert 1 - z^c \rvert \lvert K(0) \rvert \lvert K(1) \rvert > 0 .
\end{align}
Therefore, no eigenvalue $\sigma$ is found in the range $kU(0) \leq \sigma \leq k U(1)$.

\subsubsection{An infinite number of eigenvalues accumulate around $kU(0)$ and $kU(1)$} \label{sec:app_accumulation}
Next, we assume $z^c$ is located slightly below $0$, i.e., $\sigma$ is slightly smaller than $kU(0)$. In this case, we can write $\lvert z^c \rvert = (k U(0) - \sigma)\left[ U_z^{-1} + \mathcal{O}(k U(0) - \sigma) \right]$ and therefore
\begin{align}
D & = \tilde{\gamma}^-(0) \tilde{\gamma}^+(1) - \tilde{\gamma}^+(0) \tilde{\gamma}^-(1) \nonumber \\
& = C(k, \sigma) \left[ kU(0) - \sigma \right]^{1/2} \sin \left\{ \nu \log \left[ kU(0) - \sigma \right] + \vartheta(k, \sigma) \right\} ,
\end{align}
where $C$ and $\vartheta$ are continuous at $\sigma = kU(0)$. Because the argument of the sine function diverges in the limit of $\sigma \to k U(0)$, we learn that $\sigma = kU(0)$ is an accumulation point of the eigenvalues; we may find a zero of $D$ in any vicinity of $\sigma = kU(0)$. The same consideration also applies slightly above $\sigma = k U(1)$. We thus understand that an infinite number of eigenvalues are involved in $\sigma < kU(0)$ and $\sigma > kU(1)$, respectively.

\subsubsection{Eigenvalues are simple}
Let us show that all the eigenvalues are simple roots of $D$. In other words, the derivative of $D$ satisfies $D_\sigma(k, \sigma_0) \neq 0$ for every eigenvalue $\sigma = \sigma_0$. The proof is given by extending \cite{howard1961note} stability theorem. We define $G(z, c) = (U - c)^{- 1/2} A_1$ with $c \equiv \sigma / k$ and rewrite $\mathcal{T}_{k,\sigma} A_1 = 0$ as $\left[ (U-c) G_z \right]_z + \left[ - U_{zz}/2 + (N^2 - U_z^2 / 4) / (U - c) - k^2 (U - c) \right] G = 0$. Multiplying $G^\dag$ to this expression and integrating it over $0 \leq z \leq 1$, we derive an equation,
\begin{align} \label{eq:Howard_theorem}
- \left[ (U - c) G_z G^\dag \right]_{z = 0} - \int_0^1 (U - c) \lvert G_z \rvert^2 dz + \int_0^1 \left[ - \frac{U_{zz}}{2} + \frac{N^2 - U_z^2 / 4}{U - c} - k^2 (U - c) \right] \lvert G \rvert^2 dz .
\end{align}
The usual stability theorem results by setting $G(0, c) = 0$ to argue that the imaginary part of $c$ is 0. Now, we pick up an eigenvalue $\sigma_0 \in \mathbb{R}$, and accordingly $c_0 = \sigma_0 / k$, and add a small imaginary perturbation such that $c = c_0 + \ii \delta$ with $\delta > 0$. Inserting $G(z, c) = G(z, c_0) + \ii \delta G_c(z, c_0)+ \mathcal{O}(\delta^2)$ into \eqref{eq:Howard_theorem} and taking the imaginary part, we derive
\begin{align*}
\delta \left[ U(0) - c_0 \right] G_z(0, c_0) G^\dag_c (0, c_0) + \delta \int_0^1 \left[ \lvert G_z (z, c_0) \rvert^2 + \left( \frac{N^2 - U_z^2 / 4}{\lvert U - c_0 \rvert^2} + k^2 \right) \lvert G (z, c_0) \rvert^2 \right] dz = \mathcal{O}(\delta^2) ,
\end{align*}
where $G(0, c_0) = 0$ is used. Because $N^2 / U_z^2 > 1/4$ is assumed, the second term on the left-hand side is positive. Therefore, the first term cannot vanish, thus yielding $G_c(0, c_0) \neq 0$. Accordingly, the expected result,
\begin{align*}
D_\sigma(k, \sigma_0) & = A_{1\sigma} (0; k, \sigma_0) = k^{-1} \frac{d}{dc} \left\{ \left[ U(0) - c \right]^{1/2} G(0, c) \right\}_{c = c_0} \\
& = k^{-1}  G_c(0, c_0) \left[ U(0) - c_0 \right]^{1/2} \quad \left[ \because G(0, c_0) = 0 \right]  \\
& \neq 0 ,
\end{align*}
is obtained.

To summarize, the eigenvalues of $\mathsfbi{M}_k$ are defined as the simple roots of $D(k, \sigma)$ and accordingly yield simple poles in the resolvent $(\sigma \mathsfbi{I} - \mathsfbi{M}_k)^{-1}$. We can align them to write $\sigma_1^- < \sigma_2^- < \ldots < kU(0) < kU(1) < \ldots < \sigma_2^+ < \sigma_1^+$ for a positive $k$. Each $\sigma_j^-$ ($\sigma_j^+$) is a function of $k$ that represents the dispersion relation of the eigenmode traveling leftward (rightward) relative to the reference steady current.

\subsubsection{Continuous part} \label{sec:app_continuous}
As seen from \eqref{eq:solution_v12}, even when $\phi$ is everywhere finite, $\bm{v}$ diverges at a specific point $z = z^c$ if $\sigma - kU(z^c) = 0$ holds somewhere in $0 \leq z^c \leq 1$. To inspect this singularity more closely, we assume that $z^c \neq 0, 1$ and $\bm{f}$ is continuous at $z = z^c$. We then expand the solution of \eqref{eq:phi} around it taking into account \eqref{eq:gamma_critical},
\begin{align} \label{eq:phi_local}
\phi = \frac{f_2(z^c)}{k N^2} + a^- (z - z^c)^{1/2 - \ii \nu} + a^+ (z - z^c)^{1/2 + \ii \nu} + \mathcal{O}( z - z^c ) ,
\end{align}
which with \eqref{eq:solution_v12} accordingly yields $v_1 \sim (z - z^c)^{- 3/2 \pm \ii \nu}$ and $v_2 \sim (z - z^c)^{-1/2 \pm \ii \nu}$. Clearly, $\lVert \bm{v} \rVert$ is unbounded for this solution.

Different from the eigenvalue case, even though the norm is divergent, $\bm{v}(z) = (\sigma \mathsfbi{I} - \mathsfbi{M}_k)^{-1} \bm{f}$ can be computed for $z \neq z^c$ as far as \eqref{eq:phi_green} is integrable at $\zeta = z^c$. If $z^c \neq 0, 1$ and $\bm{f}$ is sufficiently smooth at $\zeta = z^c$, the improper integral can be carried out for $z \neq z^c$ by taking the Hadamard finite part. Otherwise, the integration is divergent so that $\bm{v}(z)$ is undefinable for a broad range.
% \footnote{For example, if $f_2 = \delta(z - z^c)$, the integration cannot be performed. This singularity is the origin of a growing solution for a stratified Couette flow pointed out by \cite{jose2015analytical} as equation (3.24) in their paper.}
To see the global divergence in $\bm{v}$ more specifically, we consider a limit $z^c \to 0$, i.e., $\sigma$ approaching an endpoint of the line segment, $kU(0)$. Setting $z^c \gtrsim 0$ in \eqref{eq:phi_local} and assuming the continuity of $\bm{f}$, from the boundary condition $\phi(0) = 0$, we have
\begin{align}
\frac{f_2(0)}{k N^2} + a^- (- z^c)^{1/2 - \ii \nu} + a^+ (- z^c)^{1/2 + \ii \nu} + \mathcal{O}(z^c) = 0 .
\end{align}
In order for this condition to be satisfied, the coefficients should behave as $\lvert a^\pm \rvert \propto (z^c)^{- 1/2}$ unless $f_2(0) = 0$. Accordingly, $\phi(z)$ diverges over a broad range in the limit of $z^c \to 0$. Note that this issue is not relevant to the tidally forced problem as the forcing function, $\hat{\bm{f}}_n$ in \eqref{eq:vector_and_forcing}, is identically $0$ at $z = 1$ and vanishes at $z = 0$ in the limit $z^c \to 0$, i.e., $\omega_n \to k U_0$.

In the end, we have learned that a line segment $\{ \sigma \vert \sigma = kU(z), 0 \leq z \leq 1 \}$ belongs to the spectrum of $\mathsfbi{M}_k$. Physically, this continuous part of the spectrum determines the condition for a critical level existing somewhere in the system. At this level, the wave energy becomes infinite so that the stationary solution \yo{is undefinable}.

\section{Uniform shear and stratification case} \label{sec:app_uniform_shear}
For an elementary case of uniform background shear and stratification, an exact expression is available. Specifically, we shall set $N = 1$ and $U = U_0 + (U_1 - U_0) z$ with $0 < U_1 - U_0 < 2$. For this problem, by introducing $Z \equiv (kU - \sigma) / (U_1 - U_0)$, two solutions of $\mathcal{T}_{k, \sigma} \gamma^\pm = 0$ can be represented as
\begin{align} \label{eq:gamma_Bessel}
\gamma^- = Z^{1/2} I_{- \ii \nu} (Z) \quad \text{and} \quad \gamma^+ = \ii Z^{1/2} I_{\ii \nu} (Z) ,
\end{align}
where $\nu = \sqrt{1 / (U_1 - U_0)^2 - 1 / 4}$, and $I_\mu$ is the modified Bessel function of the $\mu$th order, whose expansion at the origin is
\begin{align}
I_\mu (Z) = \sum_{m=0}^\infty \frac{1}{m! \Gamma(m + \mu + 1)} \left( \frac{Z}{2} \right)^{2m + \mu} .
\end{align}
The pair of expressions \eqref{eq:gamma_Bessel} satisfies \eqref{eq:gamma_critical} at a critical level if it exists.

The vertical structure function for the one-sided boundary value problem is
\begin{align} \label{eq:A1_Bessel}
A_1 = \ii \left (Z_1 Z \right)^{1/2} \left[ I_{- \ii \nu}(Z_1) I_{\ii \nu}(Z) - I_{\ii \nu}(Z_1)I_{- \ii \nu}(Z) \right] ,
\end{align}
where $Z_1 = (k U_1 - \sigma) / (U_1 - U_0)$ is defined. Accordingly, we also have
\begin{align}
D = \ii \left (Z_1 Z_0 \right)^{1/2} \left[ I_{- \ii \nu}(Z_1) I_{\ii \nu}(Z_0) - I_{\ii \nu}(Z_1)I_{- \ii \nu}(Z_0) \right] 
\end{align}
with $Z_0 = (k U_0 - \sigma) / (U_1 - U_0)$. Around a critical level $z = z^c = (\sigma - k U_0) / \left[ k (U_1 - U_0) \right]$, \eqref{eq:A1_Bessel} is expanded in the form \eqref{eq:A_branch}, with
\begin{align} \label{eq:alpha_Bessel}
\alpha_1 = - \frac{\ii \left( k Z_1 \right)^{1/2} I_{\ii \nu} (Z_1)}{\Gamma(- \ii \nu + 1)} \left( \frac{k}{2} \right)^{- \ii \nu} + \mathcal{O}(z - z^c) .
\end{align}
This expression is used to produce Fig.~\ref{fig:algebraic}.

\section{Asymptotic formula of a Fourier integral} \label{sec:app_asymptotic_formula}

\begin{figure}[t]
\centering\includegraphics[width=0.2\columnwidth]{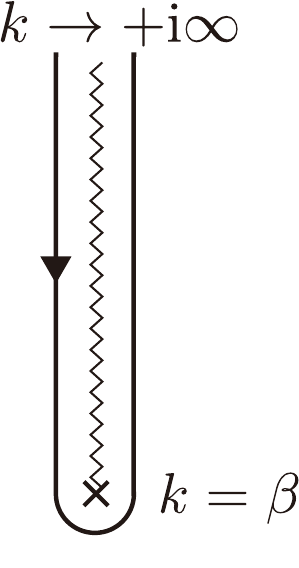}
\caption{\label{fig:hankel_contour} The integration contour of equation \eqref{eq:integral_I}.}
\end{figure}

We consider a Fourier integral,
\begin{align} \label{eq:integral_I}
\mathcal{I} = \frac{1}{2 \pi} \int_C e^{\ii k x} f(k) (k - \beta)^\alpha dk ,
\end{align}
where $x > 0$, $f(k)$ is an analytic function, and $\alpha$ and $\beta$ are complex constants. The integration contour $C$ is illustrated in Fig.~\ref{fig:hankel_contour}. We change the integration variable as $k = \beta + \ii q / x$ to write
\begin{align*}
\mathcal{I} & = \frac{\ii e^{\ii \beta x}}{2 \pi x^{\alpha + 1}} \int_C e^{-q} f \left( \beta + \frac{\ii q}{x} \right) (\ii q)^\alpha dq  \\
& = \frac{e^{\ii \beta x}}{2 \pi} \sum_{m = 0}^\infty \int_C \frac{\ii e^{-q} f^{(m)} \left( \beta \right) (\ii q)^{\alpha + m} }{m! x^{1 + \alpha + m}} dq \\
& = \frac{e^{\ii \beta x}}{2 \pi} \sum_{m = 0}^\infty \frac{\ii^{1 - \alpha - m} f^{(m)} \left( \beta \right)}{m! x^{1 + \alpha + m}} \int_C e^{-q} (-q)^{\alpha + m} dq \\
& = \sum_{m = 0}^\infty \frac{\ii^{- \alpha - m} f^{(m)} \left( \beta \right) e^{\ii \beta x}}{m! \Gamma(- \alpha - m) x^{1 + \alpha + m}} ,
\end{align*}
where Hankel's formula\footnote{e.g., \cite{bender2013advanced}, equation (6.6.22).}
\begin{equation}
\frac{1}{\Gamma(z)} = \frac{\ii}{2\pi} \int_C e^{-q} (-q)^{-z} \, dq    
\end{equation}
for the Gamma function $\Gamma(z)$ has been used.

When $\alpha = 0, 1, 2, \ldots$, we have $\mathcal{I} = 0$. Otherwise,
\begin{align}
\mathcal{I} = \frac{ f(\beta) e^{\ii \beta x}}{\ii^{\alpha} \Gamma(-\alpha)}\frac{1}{x^{1+\alpha}} + \mathcal{O}(x^{- 2 - \alpha})
\end{align}
is established for a large $x$.

\section{Derivation of the available potential energy} \label{sec:app_energy}
This section describes how to separate the total potential energy in \eqref{eq:energy_total} into a constant background part and the deviation from it, namely, the available potential energy, and makes explicit its leading-order term. First, we change the integration variable as
\begin{align} \label{eq:potential_energy}
\iint_{\epsilon h}^1 z \rho_{tot} dz dx & = \iint_{\epsilon^2 h^2 / 2}^{1/2} \rho_{tot} d \left( \frac{z^2}{2} \right) dx \nonumber \\
& = \int \left( \frac{\rho_{min}}{2} - \frac{\epsilon^2 h^2 \rho_{max}}{2} \right) dx + \iint_{\rho_{min}}^{\rho_{max}} \frac{Z^2}{2} d \rho' dx ,
\end{align}
where a function $z = Z(x, \rho', t)$ is defined as the inverse of $\rho' = \rho_{tot}(x, z, t)$ by assuming $\partial_z \rho_{tot} < 0$ everywhere holds, and $\rho_{min} \equiv \rho_{tot}(x, 1, t) = \rho_0(1)$ and $\rho_{max} \equiv \rho_{tot}(x, \epsilon h, t) = \rho_0(0)$ are understood.

Next, we define the inverse of $\rho' = \rho_0 (z)$ as $z = Z_0 (\rho')$ and introduce a variable $\eta(x, \rho', t)$ denoting the vertical displacement of an isopycnal surface as
\begin{align} \label{eq:rho_Z0_eta}
\rho' = \rho_{tot} (x, Z_0 + \eta, t) , \quad \text{i.e,} \quad  Z (x, \rho', t) = Z_0 (\rho') + \eta (x, \rho', t) .
\end{align}
Inserting the second expression into \eqref{eq:potential_energy}, we derive
\begin{align} \label{eq:potential_energy_separate}
\iint_{\epsilon h}^1 z \rho_{tot} dz dx & = \underbrace{\int \left( \frac{\rho_{min}}{2} - \frac{\epsilon^2 h^2 \rho_{max}}{2} \right) dx + \iint_{\rho_{min}}^{\rho_{max}} \left( \frac{Z_0^2}{2} + Z_0 \eta \right) d \rho' dx}_{\text{background potential energy}} \nonumber \\
& + \underbrace{\iint_{\rho_{min}}^{\rho_{max}} \frac{\eta^2}{2} d \rho' dx}_{\text{available potential energy}} .
\end{align}
The first two terms denoted by the background potential energy\footnote{More strictly, the background potential energy would be the possible minimum amount of potential energy reached by an adiabatic relocation of fluid elements. Our formulation does not fit in this definition as the third term on the right-hand side of \eqref{eq:potential_energy_separate} cannot vanish due to the non-flat bottom boundary. This issue is not essential within the scope of the current study.} do not vary with time. To prove it, because the temporal dependence of $h$ can be removed through a translational coordinate transform as inferred from \eqref{eq:first_order_boundary_condtion}, and $Z_0$ depends only on $\rho'$, it is enough to inspect the temporal variation of $\int \eta dx$. Now, we consider the amount of fluid volume with a density greater than $\hat{\rho}$, represented by
\begin{align} \label{eq:mass_0}
\mathcal{M}(\hat{\rho}) = \iint_{\epsilon h}^1 \mathcal{H}\left( \rho_{tot}(x,z,t) - \hat{\rho} \right) dz dx ,
\end{align}
where $\mathcal{H}$ is the Heaviside function. Due to the material conservation of $\rho_{tot}$, it is easy to verify that $\mathcal{M}(\hat{\rho})$ is invariant for any choice of $\hat{\rho}$. We change an integration variable of \eqref{eq:mass_0} from $z$ to $\rho'$ via $z = Z(x, \rho', t) = Z_0(\rho') + \eta(x, \rho', t)$ to write
\begin{align} \label{eq:mass}
\mathcal{M} & = - \iint_{\rho_{min}}^{\rho_{max}} \mathcal{H}\left( \rho' - \hat{\rho} \right) \frac{\partial Z}{\partial \rho'} d \rho' dx \nonumber \\
& = \int Z_0 (\hat{\rho}) dx - \epsilon \int h dx + \int \eta (x, \hat{\rho}, t) dx ,
\end{align}
where we have applied the integration by parts and used the fact that a derivative of the Heaviside function turns into the delta function and a condition $Z (x, \rho_{max}, t) = \epsilon h$. It is obvious that the first two terms in the second line of \eqref{eq:mass} are constant. Accordingly, $\int \eta(x, \rho', t) dx$ is also an invariant for an arbitrary choice of $\rho'$.

We shall expand the available potential energy, the final term in \eqref{eq:potential_energy_separate}, in terms of $\epsilon$. Let us first write $\eta = \epsilon \eta_1 + \epsilon^2 \eta_2 + \ldots$ and insert it into the first expression of \eqref{eq:rho_Z0_eta}, whose expansion is $\rho' = \rho_0 (Z_0 + \eta) + \epsilon \rho_1(x, Z_0 + \eta, t) + \mathcal{O}(\epsilon^2)$, to derive
\begin{align*}
\rho' = \rho_0(Z_0) - \epsilon N(Z_0)^2 \eta_1 + \epsilon \rho_1(x, Z_0, t) + \mathcal{O}(\epsilon^2) ,
\end{align*}
which results in
\begin{align*}
\eta_1 = \frac{\rho_1(x, Z_0, t)}{N(Z_0)^2} \quad \left[ \because \rho_0 \left( Z_0(\rho') \right) = \rho' \right] .
\end{align*}
Therefore, changing the integration variable from $\rho'$ to $z = Z_0(\rho')$, we may write the leading-order term of the available potential energy as
\begin{align}
\iint_{\rho_{min}}^{\rho_{max}} \frac{\eta^2}{2} d\rho' dx = \epsilon^2 \iint_0^1 \frac{\rho^2_1}{2 N^2} dz dx + \mathcal{O}(\epsilon^3) ,
\end{align}
to arrive at the expression \eqref{eq:available_potential_energy}.

\end{appen}

\bibliographystyle{jfm}
\bibliography{jfm}

\end{document}